\documentclass{emulateapj}
\usepackage{apjfonts}

\shortauthors{Levine et al.}

\begin{document}

\title{LOW MASS STARS AND BROWN DWARFS IN NGC 2024: CONSTRAINTS ON THE
SUBSTELLAR MASS FUNCTION}

\author{Joanna L. Levine\altaffilmark{1}, Aaron Steinhauer\altaffilmark{2}, 
Richard J. Elston\altaffilmark{3}, and Elizabeth A. Lada,}
\affil{Department of Astronomy, University of Florida,
    Gainesville, FL 32611-2055}
\altaffiltext{1}{email: levine@astro.ufl.edu}
\altaffiltext{2}{Current Address: Department of Physics and Astronomy, SUNY Geneseo, One
College Circle, Geneseo, NY 14454}
\altaffiltext{3}{Deceased}

\slugcomment{Accepted for publication in the Astrophysical Journal}

\begin{abstract} 

We present results from a near-infrared spectroscopic study of candidate brown dwarfs and low mass stars in
the young cluster NGC 2024.   Using FLAMINGOS on the KPNO 2.1m and 4m telescopes, we have obtained spectra
of $\sim$70 new members of the cluster and classified them via the prominent $J$ and $H$ band water absorption
features.  Derived spectral types range from $\sim$M1 to later than M8 with typical classification
errors of 0.5-1 subclasses.  By combining these spectral types with $JHK$ photometry, we place these
objects on the H-R diagram and use pre-main sequence evolutionary models to infer masses and
ages.  The mean age for this low mass population of NGC 2024 is 0.5 Myr and derived masses range from
$\sim$0.7-0.02 $M_\odot$ with 23 objects falling below the hydrogen-burning limit.   The logarithmic mass
function rises to a peak at $\sim$0.2 $M_\odot$ before turning over and declining into the substellar
regime.  There is a possible secondary peak at $\sim$0.035 $M_\odot$ however the errors are also consistent
with a flat IMF in this region.  The ratio of brown dwarfs to stars is similar to that found in the
Trapezium but roughly twice the ratio found in IC 348, leading us to conclude that the substellar IMF
in young clusters may be dependent on the local star forming environment.

\end{abstract}


\section{INTRODUCTION}

Brown dwarfs were first postulated four decades ago by \citet{kumar63} but their existence was not confirmed until $\sim$30
years later \citep{oppenheimer95,nakajima95,rebolo95}.  Now brown dwarfs are observed in abundance both in the Galactic
field (e.g. \citealp{kirkpatrick99,delfosse99,reid99,chabrier02}) and in young star forming regions (e.g.
\citealp{wilking99,lr00,muench01,barrado04}).  However, the origin of these objects is still unclear. Do brown dwarfs form
in a manner similar to their stellar counterparts or more akin to their planetary cousins?  What mechanism drives brown
dwarf formation and does it depend on the star forming environment?  Recently many theories of brown dwarf formation have
been proposed, including turbulent fragmentation of a molecular cloud \citep{padoan02}, ejection of protostellar embryos
\citep{reipurthclarke01}, gravitational disk instabilities \citep{pickett00}, and photo-erosion of prestellar cores
\citep{wz04}.  In depth studies of brown dwarf properties, including their spatial distribution, disk frequency, and
characteristics of the substellar initial mass function (IMF) are needed to distinguish between these formation scenarios.

The youngest star clusters (e.g. $\tau < $ 5 Myr) are ideal laboratories for the study of brown dwarfs.  Theoretical models predict that the
luminosities of contracting, pre-main sequence (PMS) brown dwarfs will be 2-3 magnitudes brighter than field brown dwarfs of
the same mass \citep{burrows97,burrows01}, making even the lowest mass objects detectable at distances out to the
nearest star forming regions (150-500 pc).  The high spatial density in most clusters ensures efficient acquisition
of a statistically significant sample of sources.  Finally, the youngest clusters have yet to undergo significant
dynamical evolution thus allowing for the direct observation of low mass stars and brown dwarfs in their natal
environments.  In other words, the observed mass distribution of a young cluster {\it is} its IMF.

In this paper we present results from FLAMINGOS photometry and spectroscopy of the young cluster NGC 2024,
including a new age and IMF for the cluster. NGC 2024 is a young ($<$1 Myr) HII region which remains deeply
embedded in its parent molecular cloud (L1630, also known as Orion B).  Distance estimates for the Orion B
cloud range from 360-480 pc \citep{at82,brown94}, with the distance to the subgroup containing NGC 2024
estimated at 415pc \citep{at82}.   Infrared photometric studies of low mass objects in the region
indicate that the star formation efficiency is high (20-40\%) and the majority of detected objects
exhibit near-infrared excess emission indicative of hot circumstellar material \citep{lada91,comeron96,hll00}.  
The proximity, extreme youth, and indicators of active star formation in NGC 2024 combine to make this an ideal region
to study brown dwarf formation.

\section{OBSERVATIONS AND DATA REDUCTION}

All imaging and spectroscopic data included in this work were taken using the Florida Multi-Object
Imaging Near-Infrared Grism Observational Spectrometer (FLAMINGOS, \citealt{flamingos}) mounted on
the Kitt Peak National Observatory (KPNO) 2.1m and 4m telescopes as part of the FLAMINGOS giant
molecular cloud survey\footnote{NOAO Survey Program 2000B-0028: {\it Towards a Complete
Near-Infrared Spectroscopic Survey of Giant Molecular Clouds} (PI: E. Lada)}.  FLAMINGOS employs a
2048$\times$2048 HgCdTe HAWAII-2 infrared array with 18 micron pixels. On the KPNO 4m telescope
this corresponds to a plate scale of 0.318\arcsec/pixel and a 10.8\arcmin$\times$10.8\arcmin field
of view.  On the KPNO 2.1m telescope, the plate scale is 0.608\arcsec/pixel, yielding a field of
view of 20.5\arcmin $\times$ 20.5\arcmin. The details of both the imaging and spectral data
acquisition and reduction are discussed in the following sections.

\subsection{FLAMINGOS Imaging and Photometry}\label{phot}   

$J$, $H$, and $K$-band images of NGC 2024 were obtained on 2001 November 19 using FLAMINGOS on the
KPNO 4m telescope.  The data were taken using a 16-point dither pattern with individual exposure
times of 60s for $J$ and $H$ and 30s for $K$, yielding total exposure times of 16 minutes in $J$ and $H$
and 8 minutes in $K$.  Typical seeing at all wavelengths was 1\farcs1-1\farcs2 FWHM.

The imaging data were reduced and photometered using the FLAMINGOS data reduction and analysis pipelines within IRAF. Briefly, the data
reduction pipeline includes linearization of all data, dark subtraction and division by a normalized flat field, a two-pass sky
subtraction routine, removal of bad pixels and finally image resampling and median combination of individual frames.  The analysis
pipeline incorporates source detection and extraction, aperture and point spread function (PSF) photometry with a 2nd order variable PSF,
astrometry, photometric calibration against 2MASS, and the combination of multi-wavelenth data into a single catalog.  It should be noted
that some small misalignments and/or fabrication errors in the FLAMINGOS optical system cause some noticeable PSF degredation which is
more pronounced at larger field positions.  In addition, images centered on NGC 2024 have a strong nebular contribution to the background
which is difficult to remove with simple aperture photometry. PSF photometry is better able to handle these variations thus we present
only PSF photometry for the remainder of this work. 

Photometric quality was assessed using a direct comparison between FLAMINGOS photometry and the 2MASS database.  The
1 $\sigma$ scatter in each magnitude bin was calculated down to the 2MASS completeness limits of J$<$15.8, H$<$15.1
and K$_s<$14.3 \citep{2mass}, and the results were combined using a weighted average with the weights dependent on
the number of sources in each bin.  In regions with little to no nebular emission, we estimate the bulk of our
photometry is accurate to within 0.03 magnitudes.   The 5$\sigma$ detection limits in these regions are J=20.1,
H=19.5, and $K$=18.5.  For regions with large amounts of nebula, as in the center of NGC 2024, the scatter with
respect to 2MASS is much larger ($\sim$0.15 magnitudes) than that expected from purely photometric noise.  A similar
effect has been noted by other authors studying young clusters with significant nebular emission (e.g.
\citealp{muench03} in IC 348 and \citealp{muench02} and \citealp{sles04} in the Trapezium) and is usually attributed
to the large size of the 2MASS pixels ($\sim$2.0\arcsec), intrinsic variability of young objects, and variations in
aperture size coupled with the strong nebular background. The photometric scatter with respect to 2MASS was also
larger for objects on the edge of detector where the data rapidly degrade due to a delamination of the engineering
array.  For objects in this region (generally non-nebular), we used photometry taken from the FLAMINGOS/KPNO 2.1m
Orion B imaging survey, taken on 2001 December 18 and reduced in the manner described above. Sources having 2.1m
photometry are identified in Table \ref{datatable} and estimated to have photometric accuracy within 0.04-0.05
magnitudes which is typical of FLAMINGOS fields observed during the 2001-2002 observing season (C. G.
Rom\'an-Z\'u\~niga et al. 2006, in preparation).

\subsection{Spectroscopic Sample Selection} \label{sample}

\begin{figure}
\plottwo{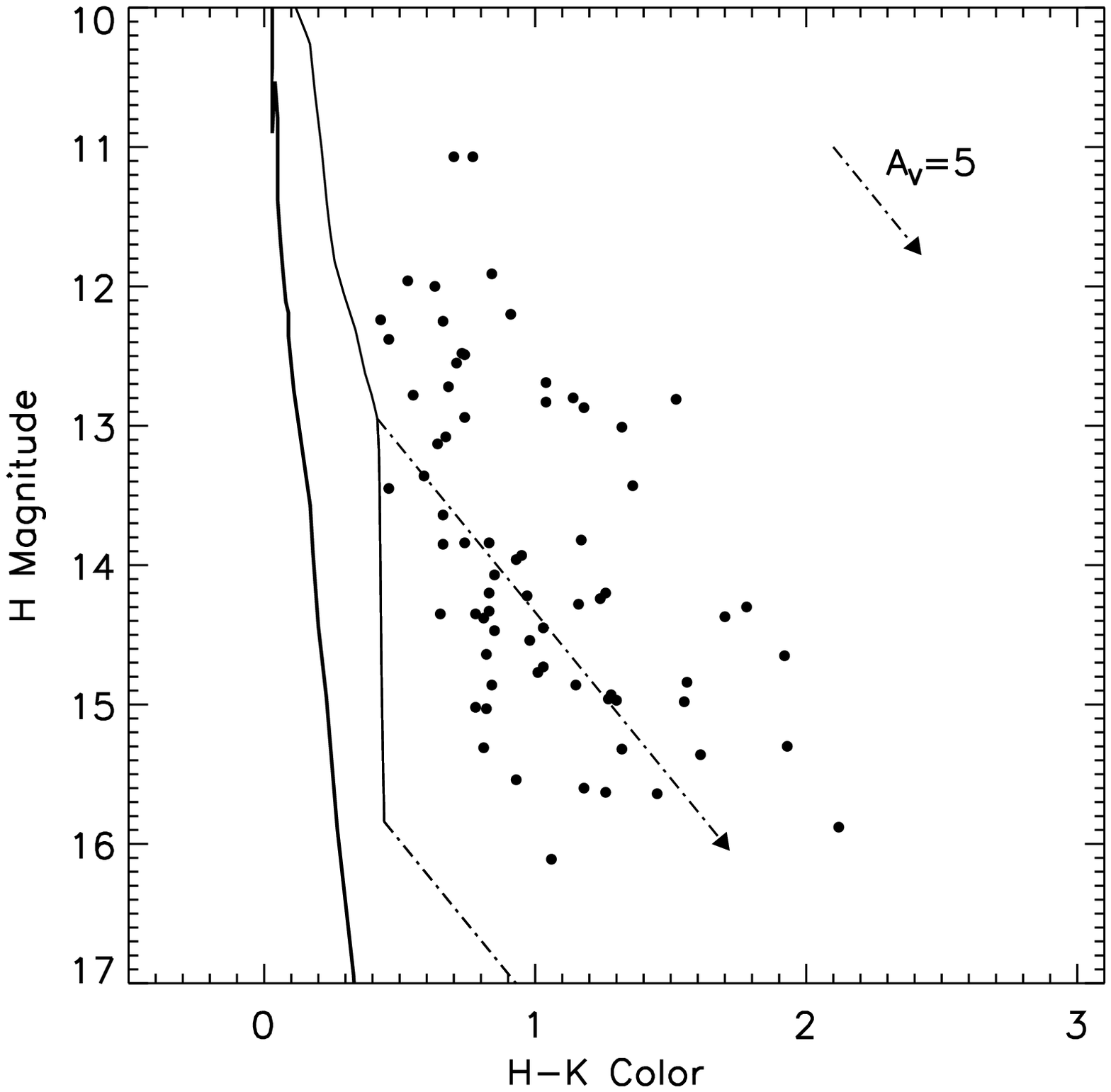}{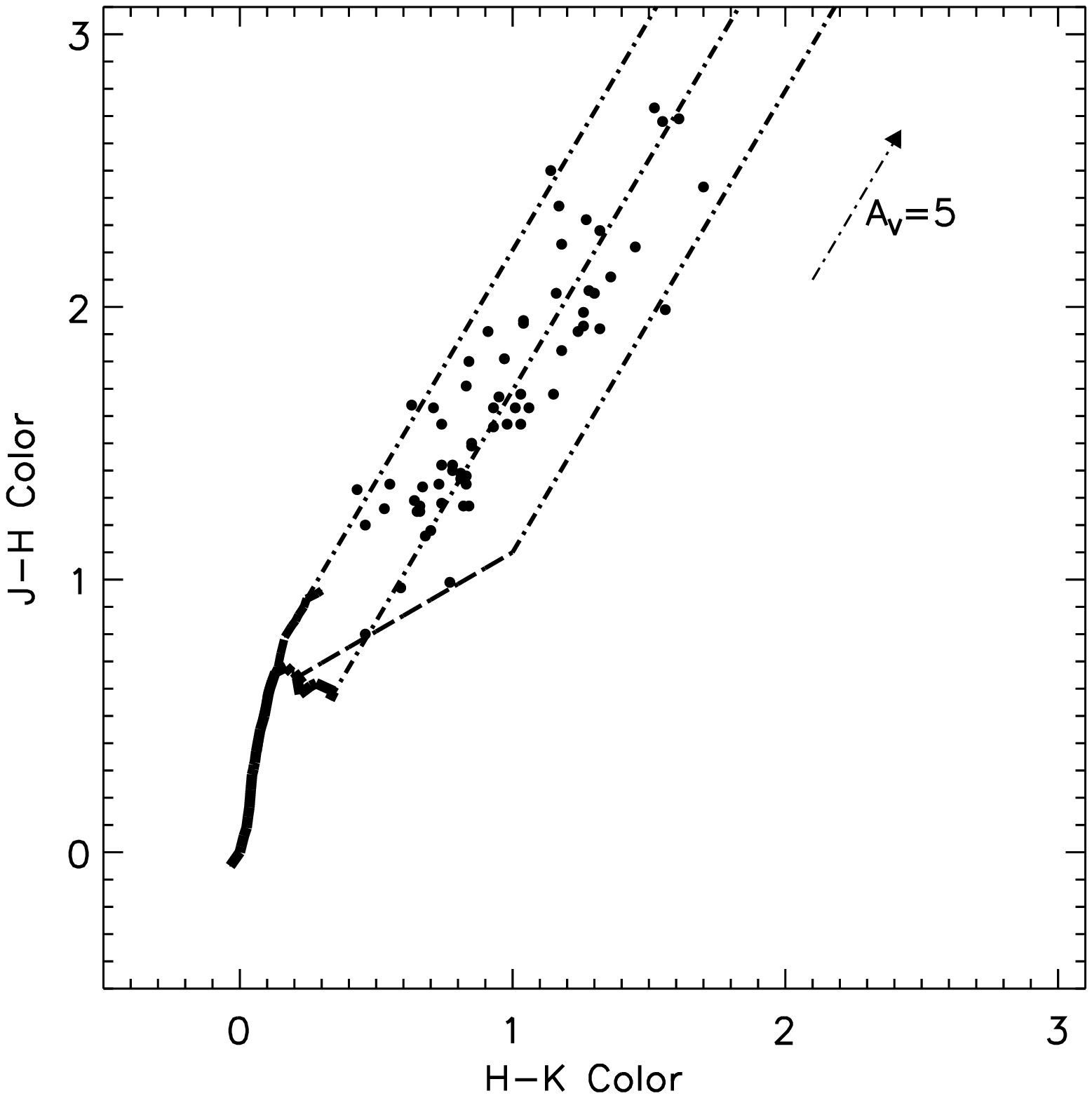}

\caption{Color-magnitude (left) and color-color (right) diagrams for all classified objects in NGC 2024.  In the CMD, the
leftmost solid line is the main sequence from \citet{bb88} and to the right is the 0.3 Myr isochrone of
\citet{dm97}.  The dot-dashed lines are reddening vectors using the extinction law of \citet{cohen81} placed
at 0.08 and 0.02 $M_\odot$.  In the color-color diagram, the solid lines are the giant colors of \citet{bb88}
coupled with a combination of \citet{bb88} dwarf colors for spectral types down to K7 and \citet{leg92},
\citet{leg96}, and \citet{dahn02} for spectral types from  M0 to M6.  The dot-dashed lines are the reddening
vectors of \citet{cohen81} and the dashed line is the classical T Tauri locus of \citet{meyer97}.  }

\label{photometry}
\end{figure}

\begin{figure}
\plotone{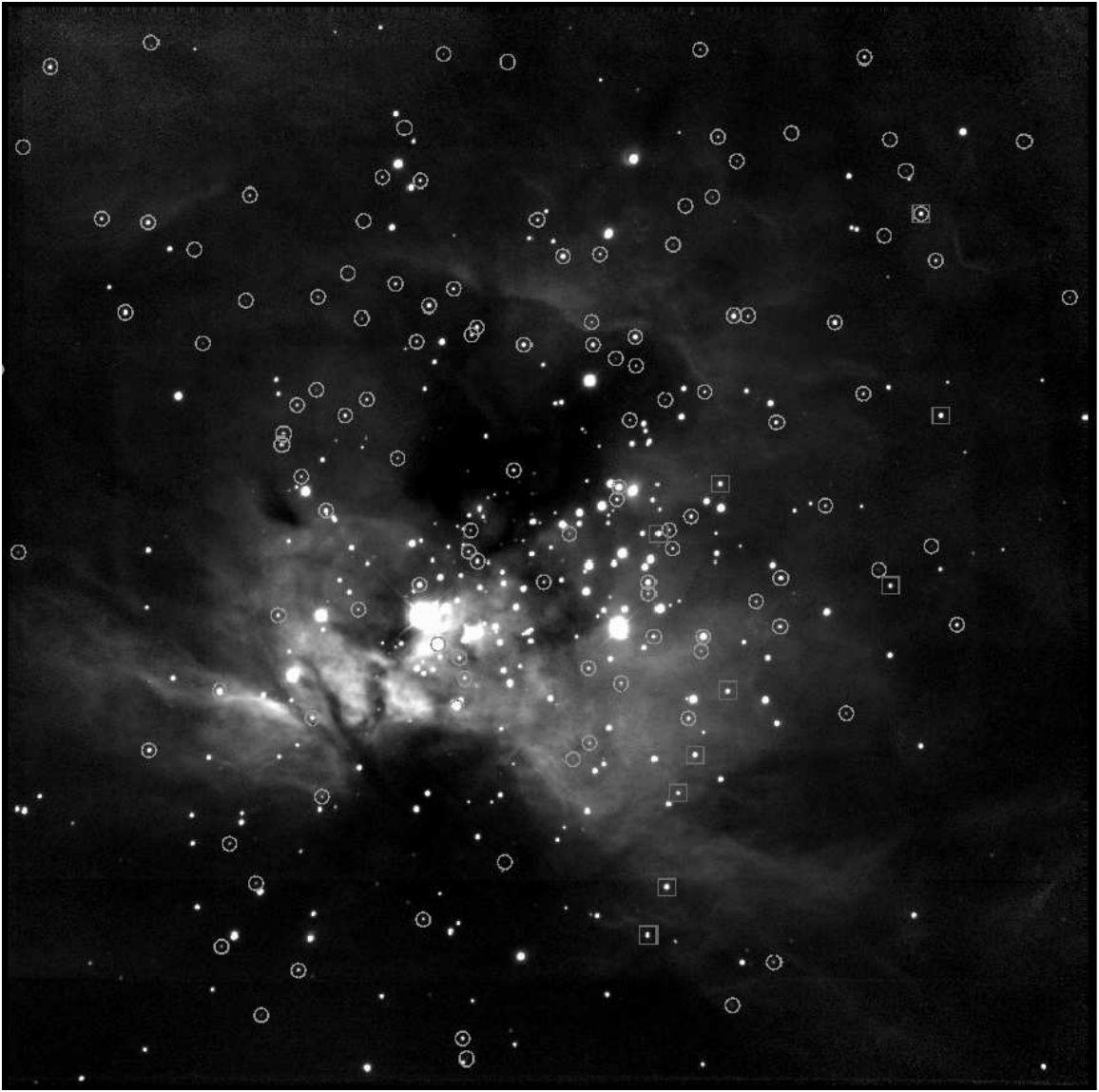}
\caption{K-band image of NGC 2024 taken with FLAMINGOS on the KPNO 4m telescope. North is up, East is to the
left, and the field is approximately 10' on a side.  Circled objects are all 4m spectroscopic targets and rectangles enclose the 2.1m targets.}
\label{image}
\end{figure}

The majority of spectroscopic targets presented in this study were observed on the KPNO 4m telescope. Target
selection for 4m spectroscopy of NGC 2024 was accomplished via the following procedure:  First, brown dwarf
candidates were selected by comparing their positions in an $H/H-K$ color-magnitude diagram (left panel of
Figure \ref{photometry}) with the 0.3 Myr isochrone of \citet{dm97} and choosing sources located below a
reddening vector extending from 0.08 $M_\odot$.  Additional young sources were selected by targeting objects
which exhibited an infrared excess (IRX) in the $J-H$ vs $H-K$ color-color diagram (right panel of Figure
\ref{photometry}).  Finally, once a slit mask was designed to target a maximum of IRX sources or brown dwarf
candidates, the remaining spaces on the mask were filled with all available objects having $K$ magnitudes
brighter than 15.0 for plates taken after December 2003 and brighter than 16.5 for plates taken earlier,
irrespective of their IRX or brown dwarf candidate status. In this manner, we targeted 120 unique sources on
five slit masks. The spatial distribution of 4m targets can be seen in Figure \ref{image}.   

A handful of brighter sources ($K<$13.0) in the region depicted in Figure \ref{image} were targeted for
observation on the 2.1m telescope as part of a larger survey to characterize the stellar population of NGC
2024.  These sources (indicated with squares in the figure) were selected solely on the basis of
their IRX status and $K$ magnitudes.  Although this sample is likely dominated by earlier spectral types, we
include these sources since any classifiable M stars in the set will aid in determining the age of the low
mass cluster population.  

It should be noted that the photometric catalog luminosity functions turn over at 18.75, 18.5, and 17.5 mag
for $J$, $H$, and $K$ respectively.  These turnovers can be interpreted as representative completeness limits, thus
we are confident that our photometry is complete well beyond the $K$=16.5 targeted limit of the spectroscopic survey.

\subsection{FLAMINGOS Spectroscopy} 
\subsubsection{4m Data Acquisition} 

FLAMINGOS spectra of the sources selected above were taken through five different slit masks using the
KPNO 4m telescope on the nights of 2003 January 19, 2003 December 06, 2003 December 10 and 2004 December
01.  Each slitlet had a width of 3 pixels (0\farcs95) resulting in a resolving power of R $\sim$ 1300. 
Sources were observed using the $JH$ filter (0.9-1.8 $\mu$m) coupled with the $JH$ grism, providing complete
spectral coverage in both the $J$ and $H$ bands simultaneously. 

For each mask, 300s exposures were taken in sets of four, dithering between two positions on the chip in
a standard ABBA pattern.  The separation between positions A and B was 4\arcsec along the long dimension
of the slitlets.  Total exposure times for each slit mask ranged from 40-65 minutes and are detailed in
Table \ref{obstable}. Short exposure, long slit spectra of a nearby G dwarf were taken immediately
following the science observations to correct for telluric absorption.  Quartz lamp flat fields and HeNeAr
arc lamp spectra for wavelength calibration were also obtained for all targets.

\subsubsection{2.1m Data Acquisition} 

Spectra of the brighter targets were obtained with FLAMINGOS on the KPNO 2.1m telescope on the night of 2003
November 29.  Each slitlet had a width of 3 pixels (1\farcs82) again resulting in a spectral resolution of R
$\sim$ 1300.  All sources were observed using the combination of the $JH$ filter and the $JH$ grism with 300s
exposures taken in sets of 4, yielding a total exposure time of 70 minutes.  Flat fields were taken using an
illuminated white screen mounted on the inside of the telescope dome and wavelength calibration was
determined using the atmospheric OH emission lines intrinsic to all NIR spectra.  Finally, a nearby G0 star
was observed to correct for telluric absorption.

\subsubsection{Data Reduction}

All FLAMINGOS spectra taken for this study (both 2.1m and 4m data)  were reduced using a combination of
standard IRAF procedures and custom FLAMINGOS routines.  We determined that a linearization correction was
not necessary as the correction was $<$1\% of the raw flux levels for usable data and the high order
correction needed would likely introduce larger errors.  Consequently, the raw data were first dark
subtracted and subsequently divided by a normalized flat field created by averaging a number of dark
subtracted quartz flats.  A pairwise subtraction of adjacent exposures was then employed to remove
background sky emission.  The sky subtracted images were aligned and combined using {\it imcombine} to
create a single image containing all spectra for each slit mask or standard star.

Long slit spectra were extracted in a straightforward manner, using the IRAF task {\it apall}.  The dispersion
correction for the long slit data was derived from similarly extracted arc lamp spectra as the background OH
emission lines were not bright enough due to the short exposure times used for the standards.  Extraction of
the multi-slit data was accomplished by first cutting a two-dimensional image of each slitlet from the final
combined image described above.  Each aperture was then traced and extracted separately using {\it apall}.  The
dispersion solutions for multi-slit data were derived using background OH emission lines local to each
slitlet.  Once the wavelength calibration was applied target spectra were divided by the G dwarf spectrum to
correct for atmospheric absorption.  Features introduced by the standard star division were removed by
multiplying the resultant spectra by the solar spectrum.  Finally, all corrected spectra were smoothed to a
resolution R $\sim$ 500 and normalized so that their flux values at 1.68 $\mu$m were unity.  Note that because
we are using a visual classification process based on broad absorption profiles (\S \ref{typing}) we are not
losing any critical information by lowering the resolution -- rather, the increase in S/N makes the overall
shape easier to discern, facillitating spectral classification.

In this manner, we were able to extract 111 individual spectra from the 4m data and 11 spectra from the 2.1m
data set.  Sources which were not extracted were either too faint to allow for an accurate aperture trace or
located at the top or bottom edge of the detector where the quality of the flat field rapidly degraded.

\section{SPECTRAL CLASSIFICATION}
\subsection{Spectral Types}\label{typing}

In order to properly classify objects in our NGC 2024 sample, it is necessary to compare the spectra of
program sources with spectra of objects with known spectral types.  The most prominent features in low
resolution infrared spectra of late-type objects are the narrow atomic lines of H I and Mg I, which decrease
in strength for later objects, and broad water absorption bands which become stronger as effective
temperatures decrease.  The water absorption bands are ideal for classifying young objects
\citep{wilking99,reid01,sles04} as they produce a very distinct continumm shape which becomes even more
pronounced for later spectral types, even at very low spectral resolutions.  However, steam absorption is
significantly stronger in the NIR spectra of young objects than in field dwarfs of the same optical spectral
type \citep{lucas01,mcgovern04}.  Consequently, if we use field dwarf standards to type our NGC 2024
members, the derived spectral types will be systematically too late.  Rather, we must use optically
classified young objects to make an accurate comparison.

Due to the dearth of published spectral types for late-type objects in NGC 2024, we turned to the nearby,
young IC 348 cluster in Perseus.  IC 348 has been extremely well studied by multiple authors (e.g.
\citealt{luhman03b,muench03} and refererences therin) and contains a number of known members with
established optical classifications. Further, the youth of the cluster (mean age $\sim$ 2 Myr,
\citealt{herbig98}) implies that spectral template stars taken from IC 348 membership lists will be similar
in age and surface gravity to our NGC 2024 program objects (see \S \ref{gravity} for further discussion of
surface gravity effects).  Table \ref{standtable} lists the identifications, positions, and spectral types
of the IC 348 standards as taken from \citet{luhman03b}.  FLAMINGOS spectra of these standards were obtained
and reduced as part of a parallel survey to classify new members of IC 348 \citet[hereafter L05]{luhman05}. 
In addition, we also obtained a spectrum of the young Taurus member KPNO-Tau 4 to provide a very late type
template (M9.5, \citealp{briceno02}).

Spectral types for candidate members of NGC 2024 were determined by visually comparing our program spectra with
FLAMINGOS spectra of the IC 348/Taurus standards, following the procedure described in L05.  In brief, spectra
were first forcibly ``dereddened" to yield a uniform continuum slope for all objects (including standards).   
This was accomplished by normalizing all spectra so that the flux at 1.68 $\mu$m had a value of 1.0. We then used
the IRAF {\it deredden} task to redden or deredden the spectra until the flux value at 1.32 $\mu$m matched the
flux of the standards (flux=1.21).  The points at 1.32 and 1.68 $\mu$m were chosen because they represent the
regions least affected by the stellar absorption features.  A spectral type for each object was then determined
by visually comparing the strength of the broad band water absorption features to those of the standards.  The
major advantage to this visual method of classification is that it is independent of the actual reddening of the
objects in question.

\begin{figure}
\plotone{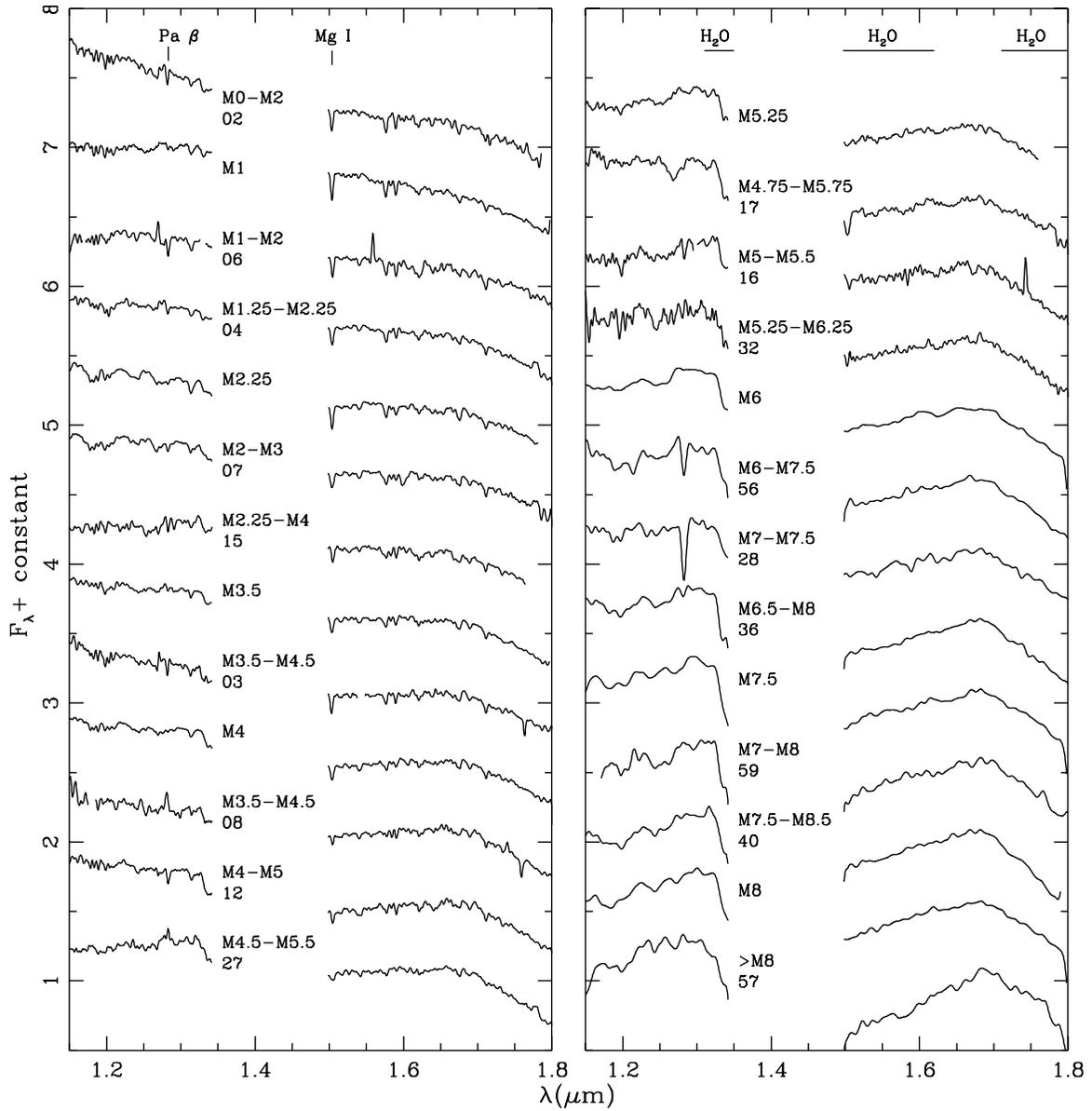}

\caption{M spectral sequence for young objects in NGC 2024 (labeled with both spectral type and ID) shown with
the IC 348 optically classified young standards (labeled with spectral type only).   Prominent spectral
features are identified at the top.  Objects having spectral types $<$M6 have been smoothed to R$\sim$500 and
objects $\ge$M6 have been smoothed to R$\sim$200 to aid in the classification process.  The central regions of
the spectra are blocked out for display purposes because in most cases the signal to noise in these regions is
very low due to the overwhelming telluric absorption. No usable information is lost from these regions.}

\label{spec}
\end{figure}

When classifying spectra, particular attention was paid to the slope and depth of the J-band fall-off at 1.35
$\mu$m, the strong absorption features in the H-band on either side of 1.68$\mu$m, and the strength of the Mg I
line.  Classifications were fine-tuned by placing all spectra in order of increasing spectral type and
adjusting the sequence to ensure that the strength of the water absorption monotonically increased with spectral
type.   We estimate this method to be quite robust, with typical errors in spectral type of $\pm$0.5-1
subclasses.  A selection of both candidate and standard spectra are shown in Figure \ref{spec} with the strongest
atomic and molecular features marked. Objects with spectral types $<$M6 are plotted with R$\sim$500 and objects
with spectral types $\ge$ M6 have been further smoothed to R$\sim$200 to aid in the classification process. 

The final classifications yielded 65 unique objects from the 4m sample with identifiable M type spectra
(ranging from M1 to $>$M8) and 2 sources with spectral types earlier than M0.  In addition, we also
extracted 4 duplicate sources which when independently classified yielded spectral types in agreement with
the original source to within 0.25 subclasses.  Of the $\sim$40 4m extracted objects which were not
classfied, 14 were filler targets with $K$ magnitudes $>$ 15.0 which we have learned are typically too faint
to classify with our current exposure times.  The remaining unclassified sources while bright at $K$, were
typically highly reddened objects with poor signal to noise in the $J$ and $H$ bands after dereddening. The 2.1m
sample yielded 4 new M type objects and one duplicate classification (which agreed with the 4m source to
within 0.5 subclasses).

\subsection{Surface Gravity}\label{gravity}

Many of the atomic and molecular features commonly found in the NIR spectra of M stars are sensitive to
surface gravity \citep{gorlova03,mcgovern04}.  This sensitivity provides a natural method for
distinguishing young sources (which have intermediate surface gravities) from relatively high
surface gravity field dwarfs or low gravity background giants.  In this section we describe the most
prominent low resolution (R$\sim$500) surface gravity indicators and discuss their implications for our data.

\begin{figure}
\plotone{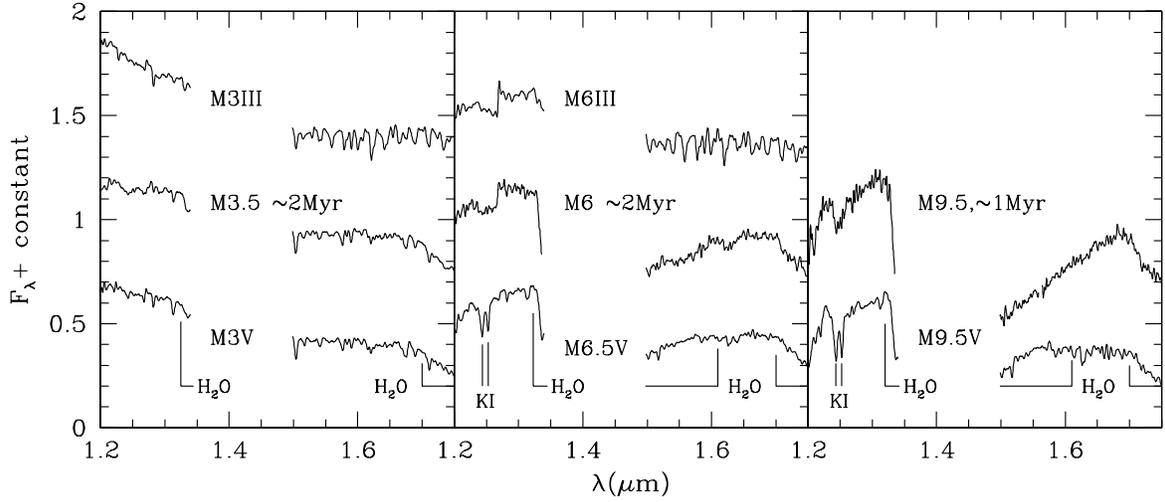}

\caption{FLAMINGOS spectra of the young objects I348-207 (M3.5), I348-298 (M6), and KPNO-Tau4 (M9.5), shown with
spectra of the field dwarfs Gl 388 (M3V), Gj 1111 (M6.5V), LHS 2065 (M9.5V) and the M Giants HD 39045 (M3III) and HD
196610 (M6III). The most prominent gravity sensitive features at R$\sim$500 are labeled.  The spectra of both giants
appear to have a much higher H-band line frequency than the young objects or field dwarfs.  In addition, water
absorption causes the younger objects to have a much more triangular H-band shape which can be used to distinguish
field stars from young cluster members.}

\label{grav}
\end{figure}

Figure \ref{grav} shows the progression of the strongest gravity sensitive features visible in our data--the broad
water absorption features in both the J and H bands and a narrow potassium doublet at 1.243/1.252 $\mu$m--as a function
of both surface gravity and spectral type.  For the M3 objects both the J and H-band H$_2$O induced fall-offs are
steepest for the young star. This effect becomes more dramatic for the M6 and M9 objects where the field dwarf
continuum profiles have broad H-band plateaus versus a distinct triangular shape for the young objects. (The same
effect was also noted by \citet{lucas01} and L05).  Looking at the M6 and M9.5 objects, the field stars have a strong
potassium doublet which is weak or absent in the lower gravity atmospheres of the young stars and giant.  Finally, it
is also apparent that the two giants have a flatter continuum profile and a much higher frequency of H-band absorption lines.  Consultation of the
literature (e.g. the low resolution infrared spectral libraries of \citet{lanconrocca92}) confirms that this is a
hallmark of giant stars and is likely caused by overtones of CO and OH as well as blended molecular lines only visible
at low surface gravities.  

Applying the above diagnostics to our data we note that the majority of sources display the distinct triangular
continuum profiles indicative of youth. We also find that the $J$-band potassium lines are absent from our spectra
therefore we are confident that there are no field dwarfs present in the sample.  Two sources (60 and 64) display enhanced
absorption in the H-band and may be background M giants.  These are noted in Table \ref{datatable}.

\subsection{Infrared Excess Emission}\label{irx}

The presence of excess flux in the near infrared is commonly taken to be an indicator of thermal
emission by warm dust in a circumstellar disk.  This type of emission can result in the weakening
or {\it veiling} of both narrow and broad-band spectral lines.  If the amount of veiling is
significant, it can affect spectral classification causing an object to appear earlier than its
true spectral type. In this section we attempt to quantify the effect of veiling on our
classification process by visually inspecting our sample and examining its infrared excess (IRX)
properties.

Visual inspection of the spectra yielded one object (source 5, M1.75) with obviously weak Mg I absorption likely
caused by veiling. In addition, excluding the possible giants (see above), 27 out of 67 objects or 40\%$\pm$9\% of
our classified M star sample exhibit an IRX as determined via a comparison of each object's expected intrinsic $H-K$
color (inferred from spectral type) with its dereddened observed $H-K$ color.  (The reader is referred to \S
\ref{ext} for an explanation of intrinsic color choice and dereddening methods.)  What fraction of IRX sources can
be expected to have significant veiling?  To answer this question we have calculated r$_k$, the $K$-band veiling
index for each source.\footnote{The veiling index r$_\lambda$ is defined as $F_{\lambda_{ex}}/F_{\lambda_*}$. 
Converting to photometric $K$-band excess yields r$_k = [(1+r_H)10^{((H-K) - (H-K)_0 - 0.065A_v)/2.5} -1]$.  Note
that because we are dereddening to intrinsic dwarf colors our values for r$_k$ are lower limits since r$_h$ is
assumed to be zero \citep{meyer97}. }   The typical errors on r$_k$ are $\pm$0.08 magnitudes, implying that we
should not place too much weight on individual values of r$_k$ which are close to zero.  However, since the primary
purpose of this r$_k$ analysis is to identify the sources with veiling strong enough to bias our spectral
classification, this error is acceptable.   As can be seen from Table \ref{datatable}, only source 38 (r$_k$=0.62)
exhibits an amount of veiling near r$_k$=0.6, the median value for Classical T Tauri stars \citep{meyer97}.  
Consequently, we note the potential bias towards an earlier spectral type for this object.

\section{H-R DIAGRAM} 

In order to construct an IMF for NGC 2024, we must determine the mass distribution of objects in our sample. In
this section we combine spectral types and infrared photometry to derive visual extinctions, effective temperatures and bolometric
luminosities for our objects.  We then place these data on the H-R diagram and use theoretical PMS evolutionary models  to infer masses
and ages for the sample.

\subsection{Extinction}\label{ext}

Because NGC 2024 is so deeply embedded in its natal cloud, there is a large and variable amount of
extinction in the region which acts to differentially redden source magnitudes.  In order to properly
estimate physical parameters and infer masses and ages, this reddening must be accounted for. The
amount of extinction towards a given source is typically derived by dereddening its broadband colors,
however, care must be taken when choosing passbands.  The redder infrared bands are less sensitive to
variations in extinction but may also be contaminated by infrared excess emission arising from a
circumstellar disk. Optical bands suffer from contamination due to UV excess emission from the
stellar photosphere.  It is generally agreed upon that bands between $R$ and $J$ are most sensitive to
extinction while minimizing the effects of excess emission ({\it cf.} \citealt{meyer97,luhman03b}). 
As we do not have reliable optical photometry for NGC 2024, we elected to use our bluest infrared
bands to derive extinction estimates for each source. 

Extinction estimates were determined by comparing our observed $J-H$ colors with the empirically determined
intrinsic M dwarf colors of \citet{leg92,leg96} and \citet{dahn02} and then converting the color excess to an $A_V$
measurement using the reddening law of \citet{cohen81}.    The choice of both the intrinsic colors and the
reddening law was based primarily on photometric system.  FLAMINGOS filters closely approximate the CIT system
(\citet{elston2003} and the FLAMINGOS web pages) thus we opted for a reddening law and intrinsic color set derived
in the same system.  We opted against using theoretical PMS colors since at young ages these are highly dependent
on model input physics.  For the one source lacking $J$-band photometry, an extinction estimate was derived using
$H-K$ colors.  It should be noted that for comparison we also estimated visual extinctions for all sources using
$H-K$ intrinsic colors and by dereddening objects to a model isochrone in both $J/J-H$ and $H/H-K$ color magnitude
diagrams. These methods yielded $A_V$ values which deviated from the $J-H$ intrinsic color estimates by as much as
1-2 magnitudes for $J/J-H$ and 3-4 magnitudes for $H/H-K$.  Effects of this deviation will be discussed in \S\S
\ref{tsect} and \ref{age}.

\begin{figure}
\plotone{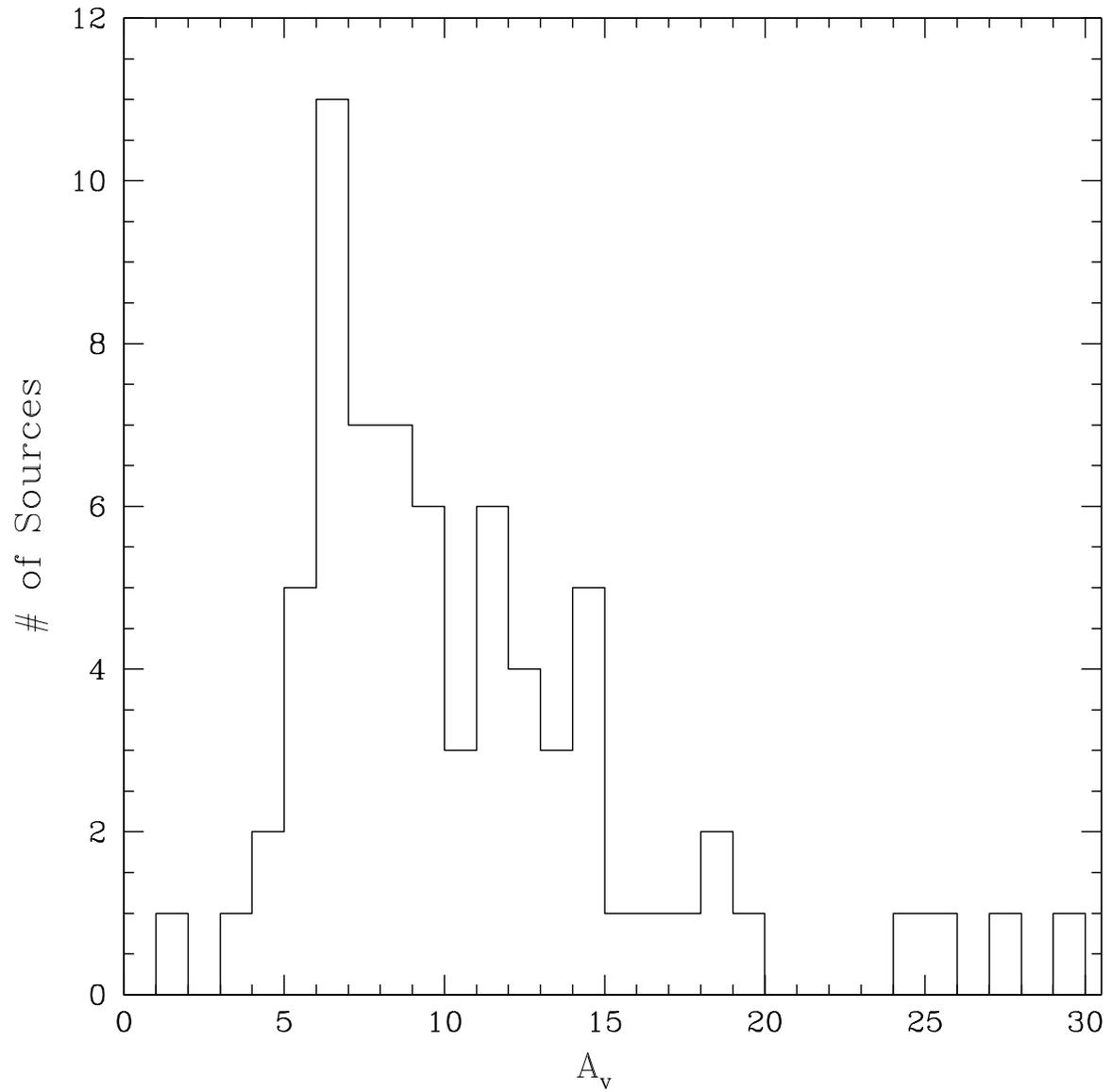}

\caption{Distribution of visual extinctions for the spectroscopic sample.  $A_V$ values were derived by
comparing observed $J-H$ colors with the intrinsic $J-H$ colors of \citet{leg92,leg96,dahn02}.}

\label{avhist}
\end{figure}

Figure \ref{avhist} shows the distribution of visual extinctions derived from $J-H$ intrinsic
colors. Values range from $\sim$1-30 visual magnitudes with a mean $A_V$ of 10.7 magnitudes.  This is in
good agreement with the survey of \citet{hll00}, who find a range of $A_V$ from roughly 0-30 visual magnitudes and a mean
$A_V$ of 10.4.

\subsection{Effective Temperatures and Bolometric Luminosities}\label{tsect}

Spectral types were converted to effective temperatures using a linear fit to the adopted temperature
scale of Luhman et al. 2003.  This temperature scale, derived from the young quadrouple system GG Tau,
falls between dwarf and giant temperature scales and is thus appropriate for the intermediate surface
gravity objects studied here.  Absolute magnitudes were calculated by dereddening K magnitudes (see
below) using the $A_V$ derived in \S \ref{ext} and applying a distance modulus of
8.09 \citep{at82}.  Bolometric magnitudes and luminosities were derived using the bolometric corrections
of \citet{leg92,leg96} and \citet{dahn02} as they were observationally determined using CIT
photometry.

While $J$-band is typically the preferred wavelength for deriving bolometric luminosities as
contaminating excess effects are minimized \citep[][see also \S \ref{ext}]{luhman99}, we have elected
to use the $K$-band since luminosities derived from $K$ magnitudes are far less sensitive to errors in
dereddening. As discussed above, photometrically derived extinction values can have errors as large as 3-4
magnitudes.  A change in $A_V$ of 3 magnitudes corresponds to nearly a magnitude of uncertainty in dereddened
$J$-band magnitudes but yields a much smaller $\Delta K$ ($<$0.3 mag).  Although $K$ magnitudes are
more sensitive to excess emission from a warm circumstellar disk, this effect is small in
log-Luminosity space (average $\Delta logL\sim$0.06 dex).  Even when combined with the $A_V$
uncertainty, the net uncertainty in $K$-derived bolometric luminosities ($\pm$ 0.17 dex) remains smaller
than the corresponding uncertainty using $J$-band to derive bolometric luminosities ($\pm$ 0.32 dex).

\subsection{Masses and Ages}\label{models}

In order to derive mass and age estimates for young sources we must place the sources on an H-R diagram
and compare them with pre-main sequence evolutionary models.  The most frequently used models for low mass stars
and high mass brown dwarfs (rather than planetary mass objects) are those of \citet[hereafter
DM97]{dm97} and \citet[hereafter BCAH98]{bcah98}.  The primary differences between these two models are
their treatment of convection (mixing-length theory for BCAH98 and full spectrum turbulence for DM97)
and the assumption of grey atmospheres in DM97 vs. non-grey in BCAH98. BCAH98 and references therein
argue that the grey atmosphere approximation is inappropriate for stars whose effective temperatures
fall below $\sim$4500-5000 K as molecules present in the atmospheres will introduce strong non-grey
effects.  There is some evidence supporting this claim as both \citet{white99} and \citet{luhman03b}
used empirical isochrones definined with low mass members of IC 348 and Taurus and the young quadrouple
system GG TAU to show that the BCAH98 models agree better with observational constraints. 
Consequently, while we present H-R diagrams using both sets of tracks, for the remaining discussion we
will focus primarily on results derived from the BCAH98 models.

\begin{figure}
\plottwo{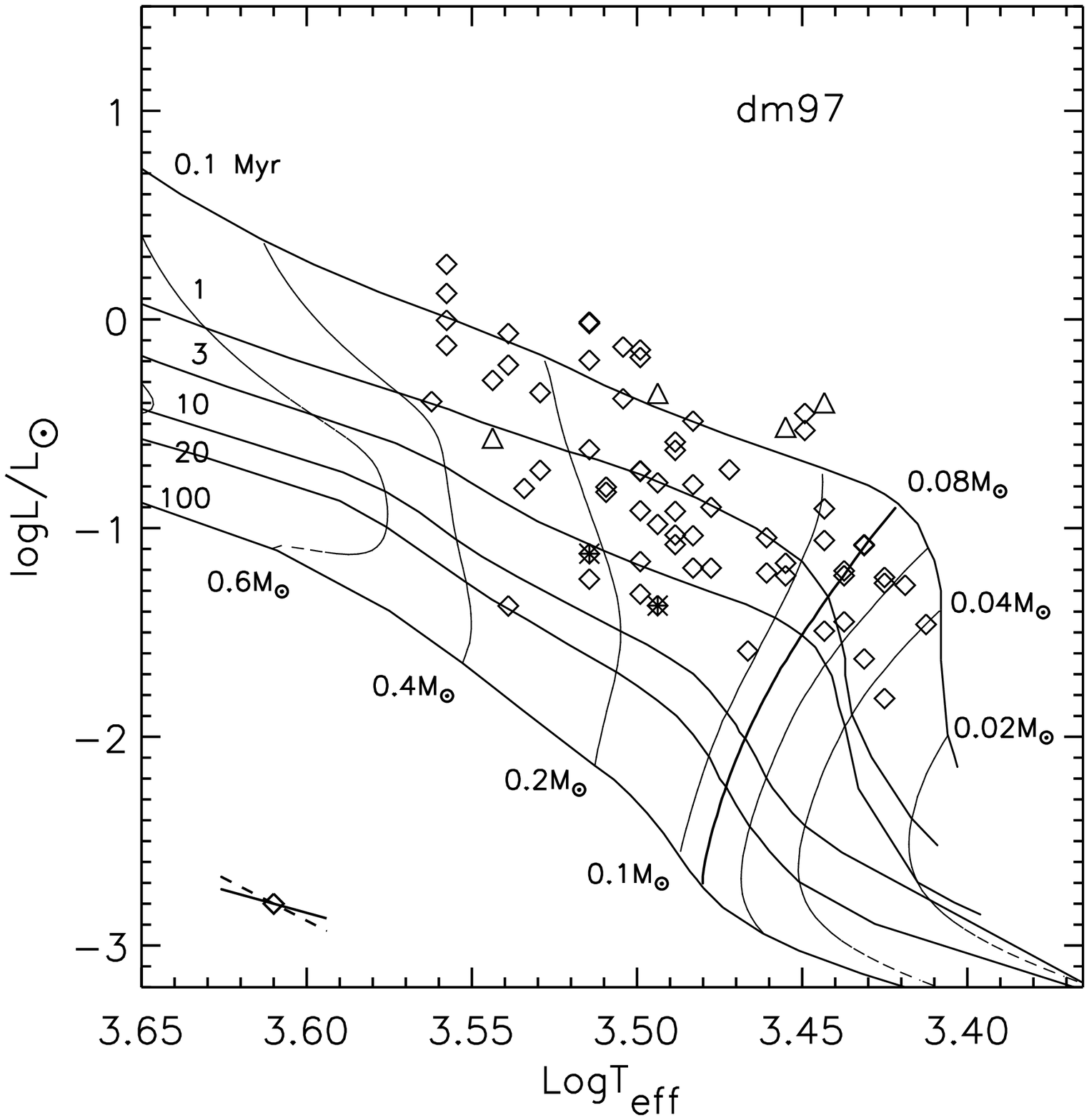}{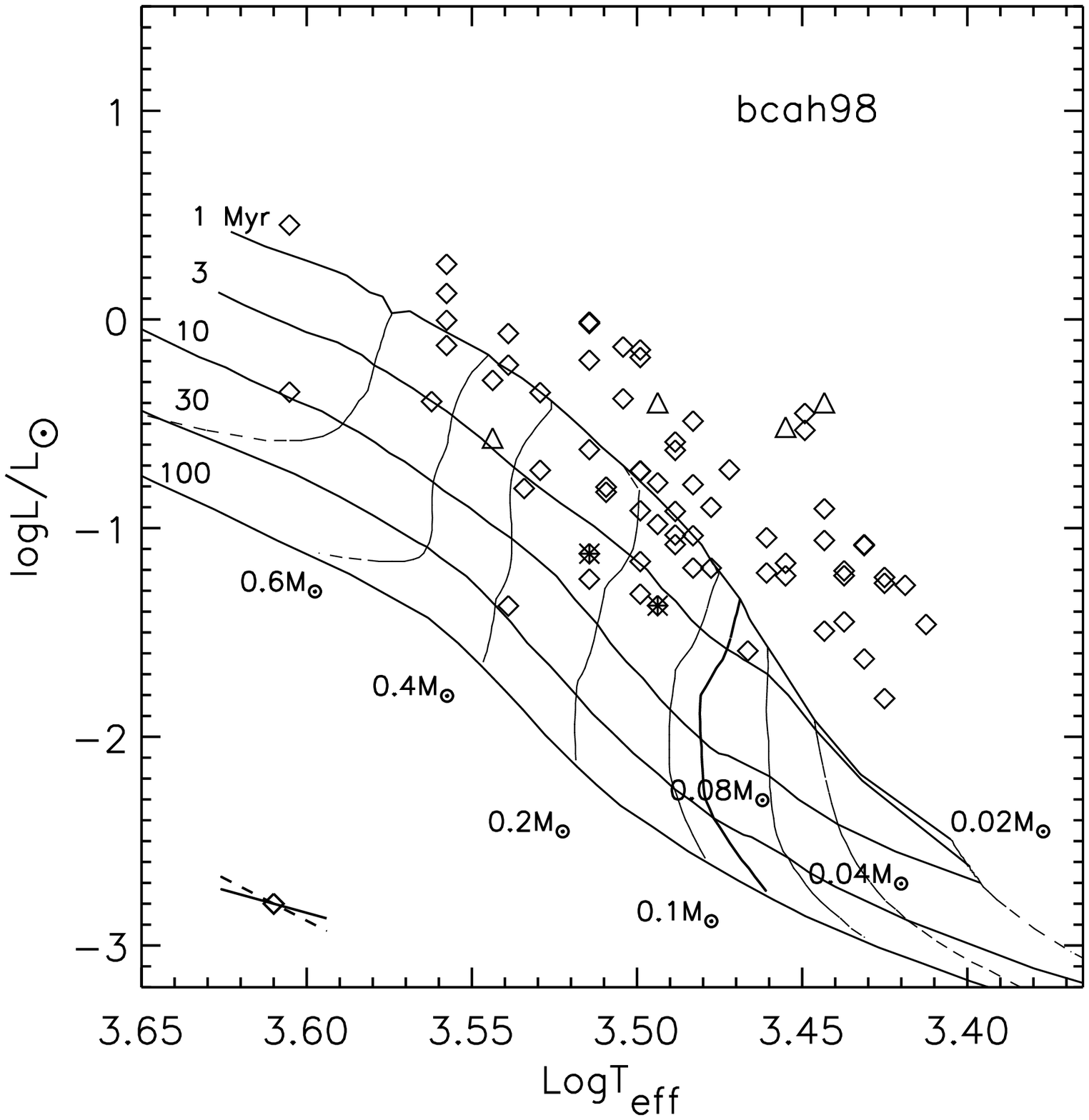}

\caption{H-R Diagrams for NGC 2024 shown with the pre-main sequence models of \citet{dm97} (left) and
\citet{bcah98} (right). The diamonds represent points with 4m spectra and the triangles are sources
classified with 2.1m spectra. Asterixes are potential background giants.  Representative error bars for an M5
object are shown. The solid line accounts for errors in derived spectral type, distance modulus, and
photometry and the dashed line incorporates an additional error of $\pm$3 magnitudes of visual extinction
(see \S \ref{ext}).}

\label{hrdiagrams}
\end{figure}

Figure \ref{hrdiagrams} shows H-R diagrams for our classified sources in NGC 2024 along with the
PMS evolutionary models of \citet{dm97} and \citet{bcah98}.  The triangular points
represent 2.1m classifications and sources with diamonds were classified using 4m spectra.  The two
asterixes represent the possible background giants (see \S \ref{grav}).  Individual
object data are tabulated in Table \ref{datatable}.  Two typical error bars for an M5 dwarf are shown in
the lower left corner.  The solid line was derived by classically propagating the measured errors in the
photometry, spectral type, and distance modulus.  In this case, the error in luminosity is dominated by
error in the distance to NGC 2024.  The dashed line incorporates an additional error of $\pm$ 3
magnitudes (corresponding to $\pm$0.2 dex) in the reddening estimate (refer to \S \ref{ext}) which
dominates the error bar and leads to a larger uncertainty in the calculated luminosity.

Mass and age estimates were derived from the BCAH98 models by interpolating between the isochrones and
mass tracks shown in Figure \ref{hrdiagrams}.  Sources falling above the youngest isochrone (1 Myr) were
assumed to have an age $<$1 Myr and were dropped down to the 1 Myr isochrone along a line of constant
effective temperature to derive a mass estimate.  In this manner we derived masses spanning a range from
0.02 to 0.72 $M_\odot$ (with 23 objects falling below 0.08 $M_\odot$) and ages ranging from $<$1 to
$\sim$30 Myr.   

\section{PROPERTIES OF NGC 2024}
\subsection{Cluster Membership}\label{mem}

Prior to drawing any conclusions regarding the age of NGC 2024 or its substellar population, it is necessary to evaluate
the membership status of sources in our sample.  In the absence of proper motion data, we must rely on other diagnostics to
determine whether objects are bona fide cluster members or foreground or background sources projected on the cluster area. 
The discussion of surface gravity effects in \S \ref{gravity} rules out foreground or background dwarf contamination in our
spectroscopic sample as there are no potassium lines present in our spectra.  In addition, NGC 2024 is deeply embedded in a
core of dense gas \citep{lada91,lada97} which will obscure background field stars, limiting the number of field
contaminants in the photometric sample.  The average column density of hydrogen in a 0.6 pc clump centered on NGC 2024 has
been estimated from C$^{18}$O emission to be $N$(H$_2$)=4.6$\times$10$^{22}$ cm$^{-2}$ \citep{aoyama01}.  Given that a
molecular hydrogen column density of 10$^{21}$ cm$^{-2}$ corresponds to 1 magnitude of visual extinction \citep{bohlin78},
background sources in this region will be viewed through 46 magnitudes of visual extinction, or $\sim$4.1 magnitudes of
$K$-band extinction.  The spectroscopic sample includes sources down to $K\simeq$15.  Background objects contaminating this
sample are seen through the cloud and thus will have unreddened magnitudes $K\leq$11.  Looking at the distribution of
sources in an off-cloud FLAMINGOS control field as well as a similar area from the 2MASS database, we estimate that there
are no more than 5 background sources with K$\leq$11. As this is a relatively insignificant contribution to the total
photometric KLF, we conclude that a background correction is unnecessary. We do note the possibility of giant contamination
for two spectroscopic sources (60 and 64) which display enhanced absorption in the H-band and we exclude these objects from further
analysis.

\subsection{Age of NGC 2024}\label{age}

Sources in the H-R diagrams in Figure \ref{hrdiagrams} do not fall along a single isochrone but rather
show a scatter in age ranging from $\ll$1 Myr to $\sim$30 Myr, irrespective of the PMS models used. This
type of width in the evolutionary sequence of young clusters is common and is usually attributed to a
variety of effects including: real age differences between sources, errors in luminosity derived from
uncertainties in the derived reddening, photometric uncertainty (these effects are represented by the
error bars in the figure), as well as distance variations between sources, variability due to accretion
and rotation of young objects, and unresolved binaries.   The median age of the entire sample however
should be representative of the median age of the cluster population in the mass range detected here.  

Using the models of BCAH98 the majority of sources fall above the 1 Myr isochrone.  Consequently, the
median age of the cluster can only be constrained to $<$1 Myr.  However, the DM97 models extend to
younger ages than those of BCAH98.  Even though we have elected to place more weight on results derived
with the BCAH98 models (see \S \ref{models}), the DM97 models provide us with additional information on
the age of the cluster population as well as a means to compare our results with previous surveys of NGC
2024 and other regions where authors have used the DM97 models to derive an age.  Using the DM97 models we
derive a median age of 0.5 Myr.  If we factor in errors in the distance modulus (8.09$\pm$0.17,
\citealp{at82}), this leads to a possible age range of 0.4-0.6 Myr.  Including a 3 magnitude shift in
reddening (refer to \S \ref{ext}) yields a larger range of 0.2-0.9 Myr.  All of these results remain
consistent with the age derived from the BCAH98 models, placing NGC 2024 at $<$1 Myr.

Our derived age of 0.5 Myr for NGC 2024 is in good agreement with ages found by previous surveys of NGC
2024.  Both \citet[][hereafter M96]{meyer96} and \citet{ali98} used infrared photometry and spectroscopy
with the models of \citet{dm97} to derive mean ages of 0.3 and 0.5 Myr respectively.   Further, M96 used a
distance modulus of 8.36 magnitudes.  Increasing the distance to the cluster acts to increase the derived
bolometric luminosity of sources, making objects appear younger.  Indeed, if we use the larger distance
modulus with the models of DM97, we derive a median age of 0.3 Myr which is in excellent agreement with
the results of M96.

A few sources in our H-R diagrams appear to have ages which deviate significantly from the median. Using either
set of PMS models there is a small, lower luminosity population with inferred ages $>$3 Myr.  In order to
attribute these low luminosities to general scatter caused by photometric errors and uncertainties introduced by
variability (generally no more than $\pm$0.2 mag at $K$), derived reddening ($\pm$0.3 mag, see above), and
distance modulus ($\pm$0.2 mag), these effects would have to combine to produce at least a 1-2 magnitude shift at
$K$.  On the other hand, it has been noted by multiple authors (e.g. \citealp{luhman03b,sles04,wilking04}) that a
circumstellar disk can act to occult the central source, resulting in an underestimate of the object's luminosity
and thus an overestimate of the object's age.  We have examined the infrared excess properties of the subsample in
question and indeed, irrespective of the model isochrones used, all but one of the objects with an inferred age
$>$3 Myr have excess flux, indicating the presence of circumstellar material.

Looking at Figure \ref{hrdiagrams}, we also notice that object age may be slightly mass dependent with the less
massive sources appearing younger.  To quantify this trend, we have divided our sample into two populations: objects
with masses lower than the median mass and objects with masses higher than the median mass (M$_{median}\simeq$0.15
M$_\odot$).  Using the BCAH98 models we are unable to detect an age difference between the low and high mass samples
as both have median ages $<$1 Myr.  However, the median ages indicated by the DM97 models are somewhat different from
one population to the other.  The low mass sample has an age of 0.3 Myr and the higher mass sample has an age of 0.9
Myr.   There are a number of possible explanations for this effect.  First, the trend may be an artifact arising from
uncertainties in the evolutionary models at very young ages and low masses.  No two sets of PMS tracks look alike in
the brown dwarf regime thus it is a distinct possibility that the apparent age segregation is caused by a problem
with the tracks.   On the other hand, the observed mass dependence could be a selection effect caused by the
intrinsic faintness of the older substellar population -- according to the BCAH98 models even the highest mass brown
dwarfs will be undetectable by our survey by the time they reach ages of 2-3 Myr and this limit becomes younger for
lower mass objects.  Finally, it is also possible that this effect is real.   If so, this may be evidence for
sequential formation as a function of mass where lower mass objects form later in the evolutionary sequence of a
young cluster.   Unfortunately, our data are not sensitive enough to distinguish between these possibilities --
deeper spectroscopic observations are needed.

\subsection{The Substellar Population}\label{bds}

\subsubsection{Spatial Distribution of Sources} 

\begin{figure}
\plotone{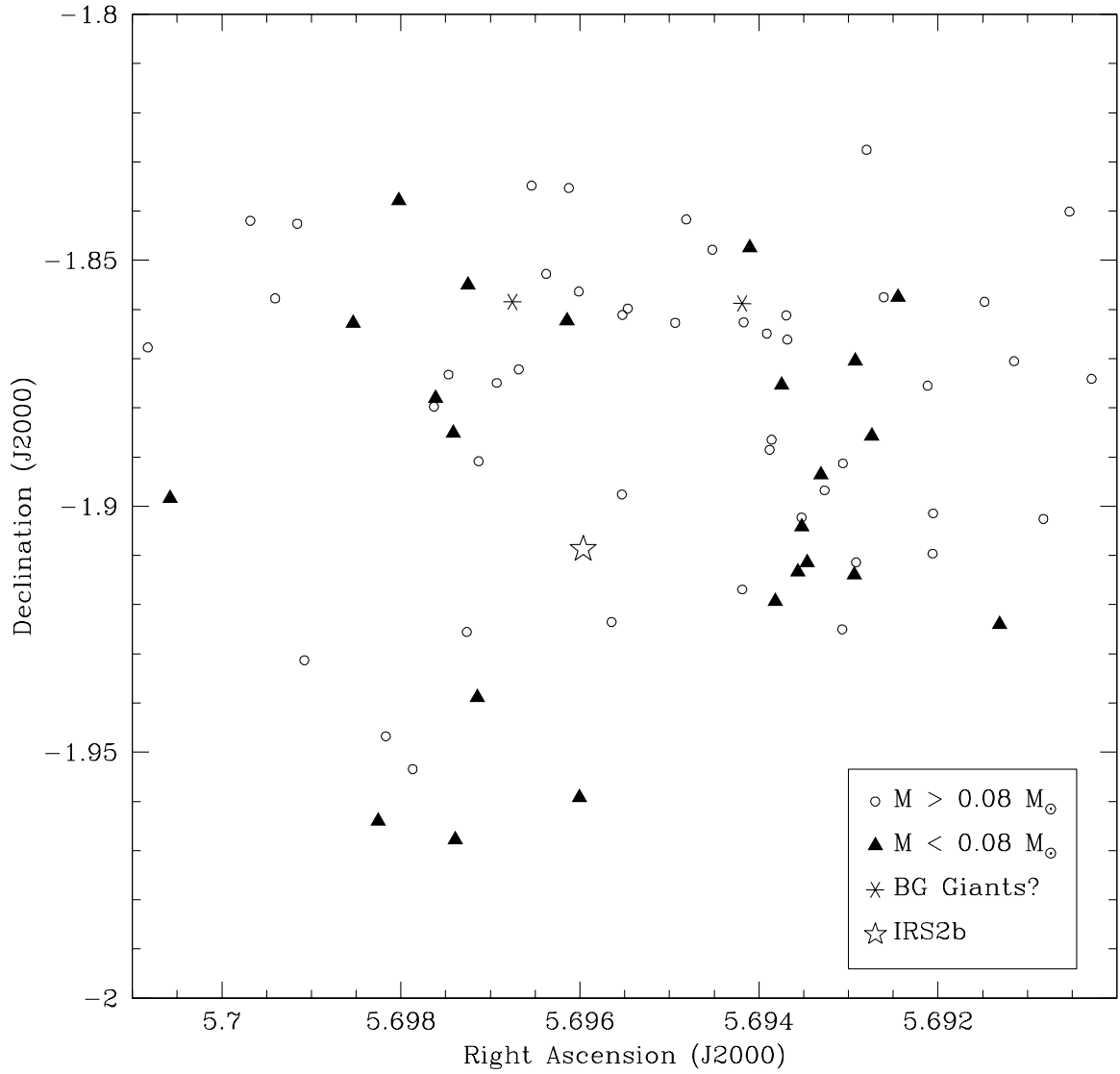}
\caption{Spatial distribution of both stellar and substellar objects in NGC 2024. The open circles represent
objects with masses M $>$ 0.08 $M_\odot$, filled triangles are objects with M $<$ 0.08$M_\odot$, asterixes are
possible background giants, and the star represents IRS2.}
\label{bddist}
\end{figure}

Figure \ref{bddist} presents the spatial distribution of all sources classified using FLAMINGOS spectra. 
Open circles are objects with M $>$ 0.08 $M_\odot$, filled triangles have masses M $<$ 0.08 $M_\odot$,
and asterixes are the possible giants.  The star at the center of the cluster represents IRS2b, the likely ionizing source for the region (see below).  It can be seen from this figure that the substellar objects are
not localized to one region but rather appear to be distributed similarly to the stellar mass
objects classified here.  It should be noted that the dearth of classified objects (either stellar or
substellar) in the center of the cluster is a selection effect caused by the high extinction in this
region blocking much of the $J$ and $H$-band flux. 

\subsubsection{Disk Frequency}

As discussed in \S \ref{irx}, the presence of an infrared excess is commonly taken to be an indicator of thermal emission from a
circumstellar disk.  Disks around brown dwarfs are of particular interest because their presence or absence has implications for the
likelihood of planet formation (planets form within circumstellar dust disks) and the formation mechanism of brown dwarfs (accretion
disks play an important role in the star formation process).  Combining our spectral classifications with $H-K$ intrinsic colors we
find 40\%$\pm$9\% of sources in our total sample have an $H-K$ color excess.  This method of selecting excess sources has been shown
by \citet{liu03} to be more sensitive to small IR excesses (as opposed to the traditional $JHK$ color-color diagrams) and is thus well
suited for investigating the disk properties of brown dwarfs, which are expected to have smaller excesses than their stellar
counterparts.  Approximately one third of the excess sources detected using the color-spectral type analysis have masses which place
them below the hydrogen-burning limit (spectral types $\ge$M6).  This yields a substellar $HK$ excess fraction for NGC 2024 of 9/23 or
39\%$\pm$15\%, where quoted errors are derived from Poisson statistics.   

Substellar excess fractions have been compiled for a number of other regions. For example, \citet{muench01} used $JHK$ color-color
diagrams to examine a set of photometrically selected brown dwarfs in the Trapezium cluster and found a substellar excess fraction of
$\sim$65\%$\pm$15\%.  In a follow-up $L'$ study of the same region, \citet{lada04} find a $K-L'$ excess fraction of 52$\pm$20\% for
their spectroscopically selected brown dwarf sample and 67\% using a $JHKL$ color-color analysis for the larger photometric sample. 
More recently, \citet{luhman05b} used the Spitzer Space Telescope to obtain mid-infrared photometry for low mass members of the IC 348
and Chamaeleon I clusters, finding that 42\%$\pm$13\% of brown dwarfs in IC 348 and 50\%$\pm$17\% of brown dwarfs in Chamaeleon
exhibit excess emission, consistent with our result for NGC 2024.

Based on the above results (placing more emphasis on studies with spectroscopic information), we can conclude that 40-50\% of brown
dwarfs are surrounded by circum(sub)stellar disks.  Note though in many cases the quoted substellar excess fractions are deemed lower
limits to the true substellar disk fraction (e.g. \citealp{lada04,luhman05b}).  This is also true for NGC 2024.  Disk modeling by
\citet{liu03} shows that the {\it maximum} expected $K$-band excess for a disk with no inner hole is 0.42 magnitudes for an M6 dwarf
and 0.31 magnitudes for an object classified as M9.  However, the $L'$ observations of \citet{liu03} are more consistent with disks
having an inner hole $R_{in}\approx(2-3)R_*$.  The $K$-band excess for these objects would be very small or undetectable using the
$H-K$ analysis we present here.   

Our choice of intrinsic colors may also lead to an underestimate of the substellar disk fraction in NGC 2024. 
We have used an intrisic color set derived from observations of field dwarfs.  Our targets are pre-main
sequence objects which have lower surface gravities than field dwarfs (e.g. \S \ref{gravity}) and thus bluer
$H-K$ intrinsic colors for a given spectral type (refer to the low surface gravity giant sequence plotted in
Figure \ref{photometry}b as compared to the dwarf sequence plotted in the same figure).  The assumption of the
redder dwarf colors will preclude objects with a $K$-band excess similar to or smaller than the difference
between PMS and dwarf colors from being counted as excess sources.  Combining this effect with the fact that
$H-K$ excess is a poor indicator of disk emission for substellar objects (see above), we conclude the true
substellar disk fraction for NGC 2024 may be significantly higher than 39\%.  This yields further weight to
the idea that the majority of brown dwarfs form through a disk accretion process similar to their stellar
counterparts.

\subsubsection{Low Mass IMF}\label{imf} 

\begin{figure} 
\plottwo{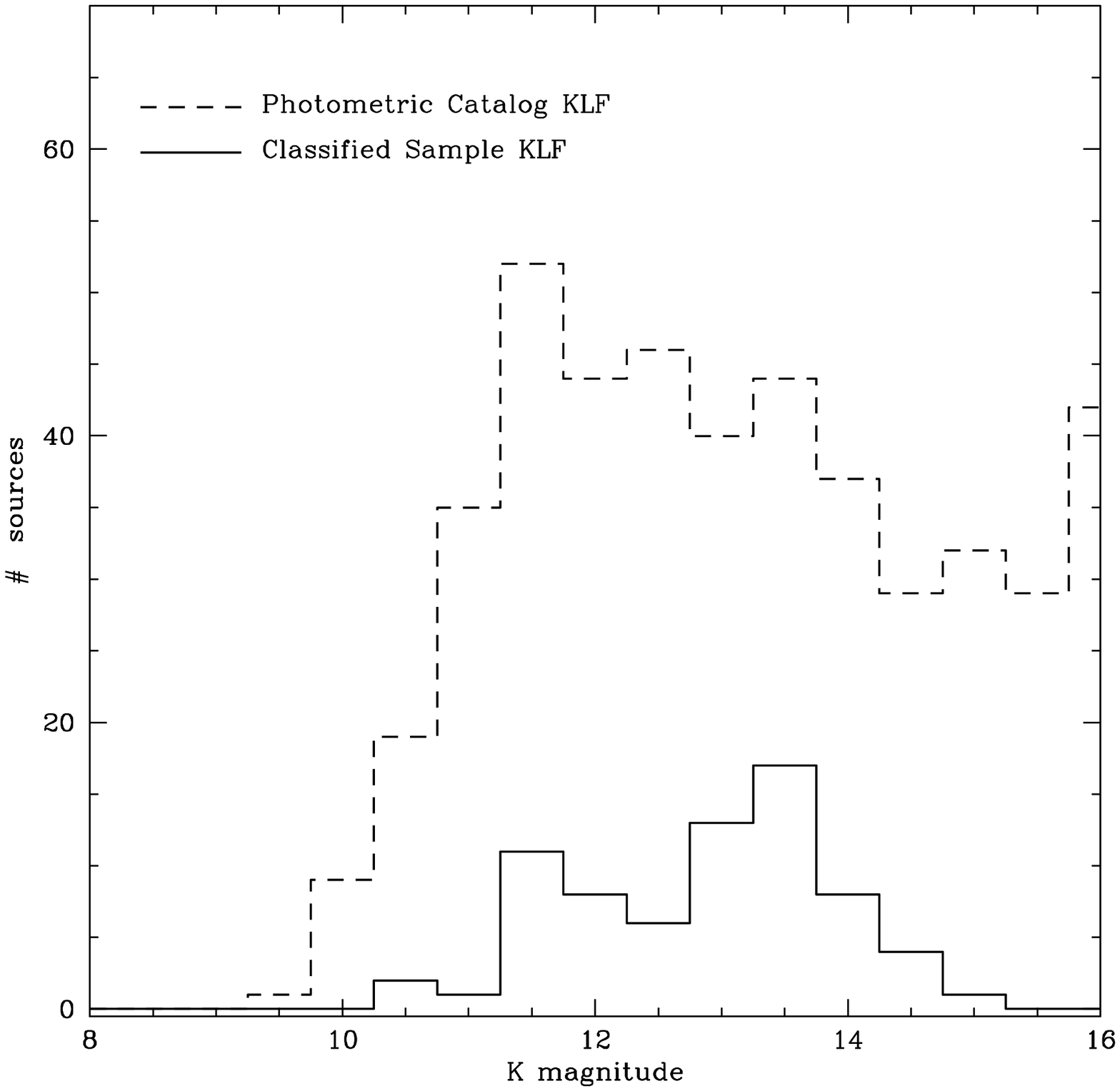}{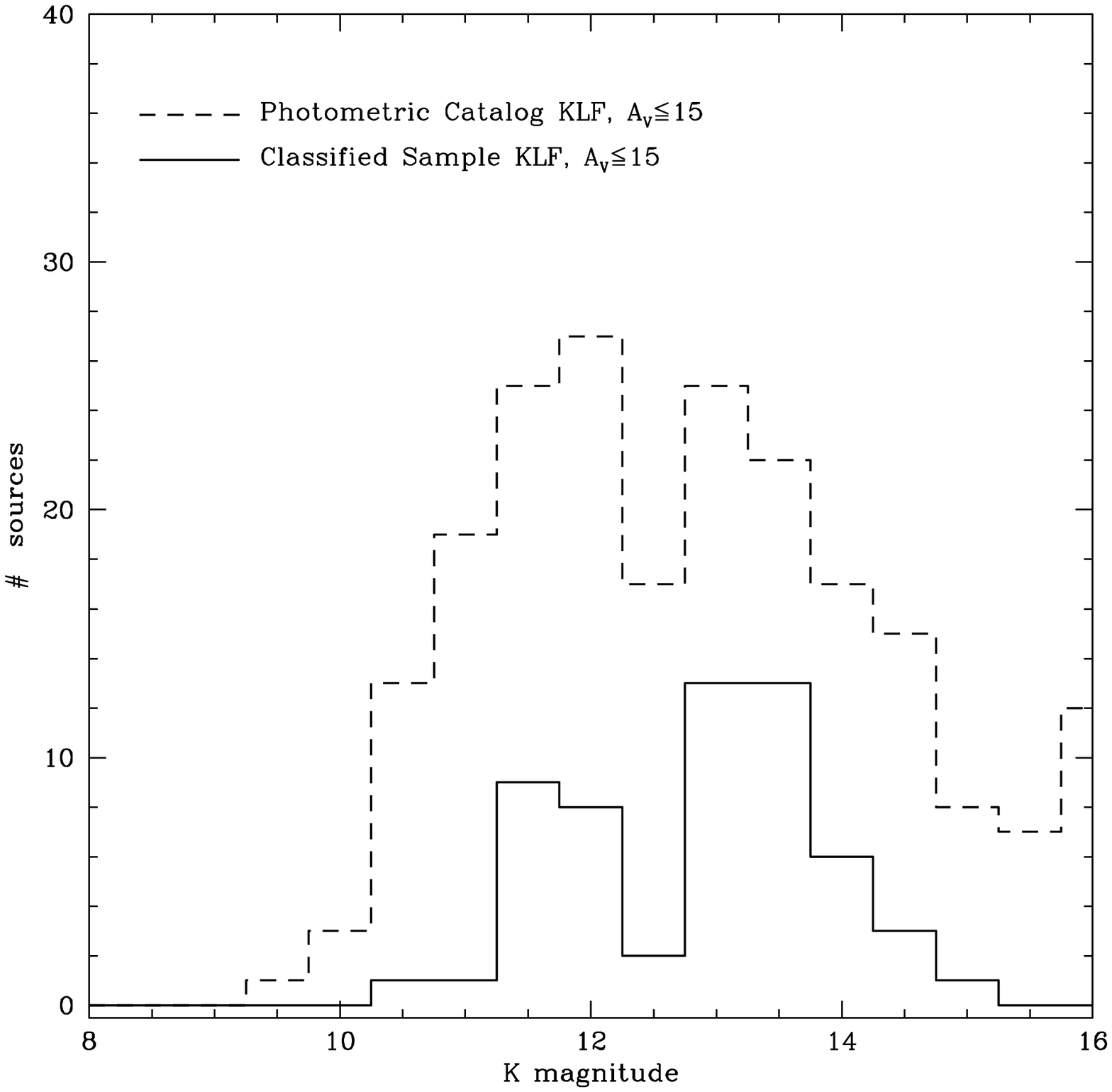} 

\caption{{\it Left:} $K$-band luminosity function for all classified objects in the spectroscopic sample (solid line) shown
with the KLF for the photometric catalog (dashed line). {\it Right:}  KLFs for both samples with an imposed extinction
limit of $A_V\leq15$.  It can be seen from the figure that in the magnitude range 11.25$<K<$14.75 the extinction-limited
spectroscopic sample is representative of the extinction-limited main cluster population.  } 

\label{finalsample} 
\end{figure}

Prior to constructing a mass function for NGC 2024, we must ensure that the subsample under consideration is representative of the
overall cluster population.  The left-hand panel of Figure \ref{finalsample} shows the uncorrected photometric luminosity function
with the KLF of the final classified spectroscopic sample.  Without placing any limits on the data, it can be seen that our
spectroscopic survey is typically only 10-20\% complete in the magnitude range from $K$=10.5-15.0.    As discussed in \S
\ref{mem}, correcting for background field stars will have little effect.  Rather, much of our incompleteness is caused by high
reddening within the molecular cloud itself (\S \ref{ext}). Imposing an extinction limit on the data yields a higher completeness
fraction and gives a more controlled sample from which we can construct an IMF.   The right-hand panel of Figure \ref{finalsample}
shows the $K$-band luminosity functions for all sources having $A_V$$\leq$15 in both the photometric catalog and the final sample
of classified objects.  Disregarding the bins on either end (as they contain only one object each), it now appears that the
spectroscopic KLF is a good representation of the total photometric KLF in the magnitude range 11.25$<$$K$$<$14.75.  Further, with
the exception of the bin centered on $K$=12.5, the completeness fraction in the same magnitude range now extends from
$\sim$25-60\% with a median value of 35\%.  Following the work of \citet{sles04},  we corrected for this incompleteness by adding
sources to each deficient magnitude bin according to the object mass distribution in that bin.  

\begin{figure}
\plotone{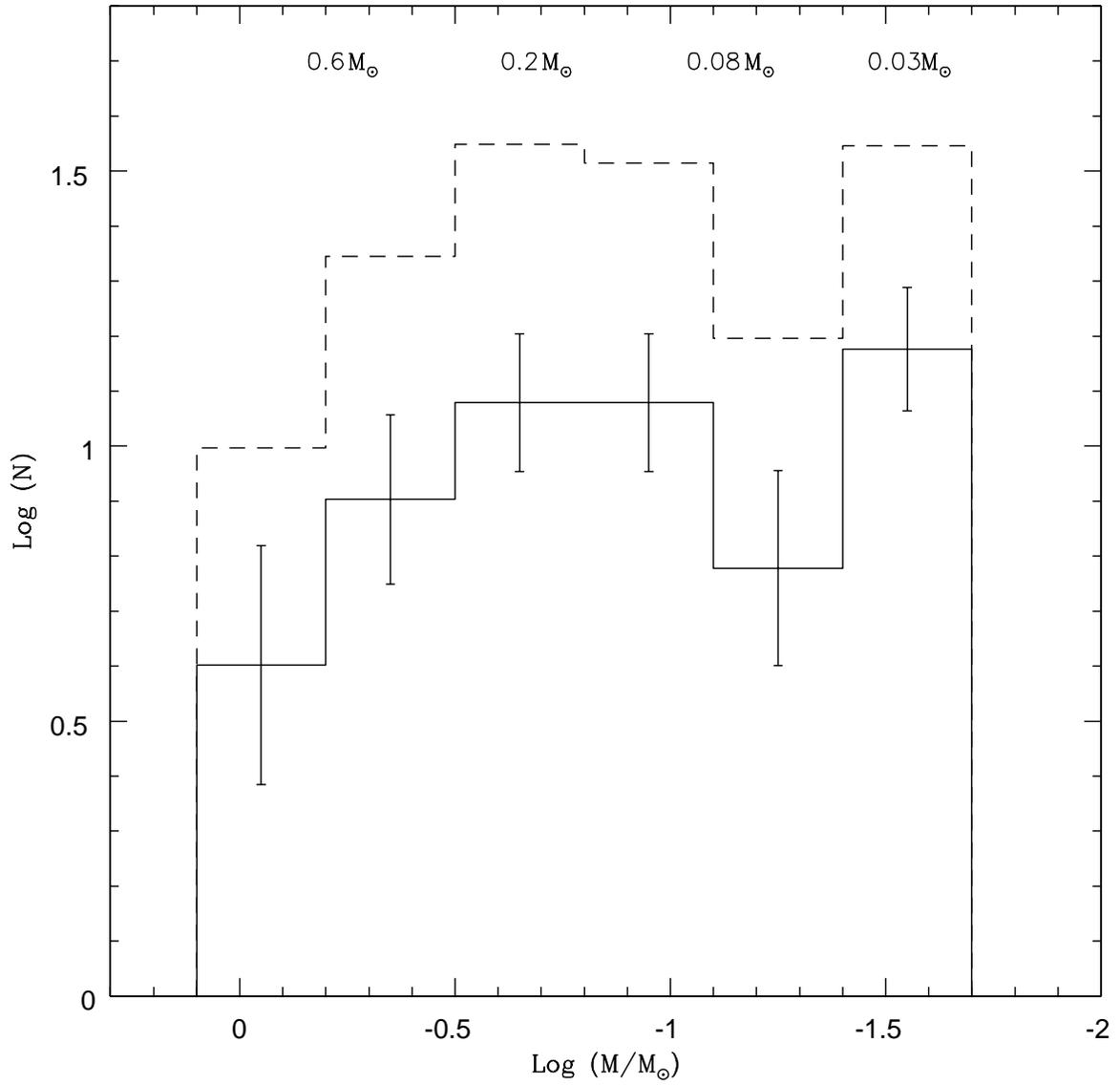}
\caption{Mass function for all classified objects in NGC 2024 whose spectra indicate that they are cluster
members with spectral types $\geq$M0.  The solid line is the uncorrected raw mass function shown with Poisson
error bars and the dashed line has been corrected for magnitude incompleteness in the range from
11.25$<K<$14.75.}
\label{massfn}
\end{figure}

Figure \ref{massfn} shows the spectroscopically derived mass function for NGC 2024.   The solid line is the mass function for all
objects with spectral types $\geq$M0, excluding the two possible giants (\S\ref{mem}).   Error bars are derived from Poisson
statistics.  The dashed line shows the IMF for the same sample corrected for the incomplete magnitude bins down to $K$=14.75.  We
estimate that for our extinction-limited sample, this corresponds to a mass completeness limit of 0.04 $M_\odot$.  The mass
function rises to a peak at $\sim$0.2 $M_\odot$ before declining across the stellar/substellar boundary.  There is an apparent
secondary peak around $\sim$0.03 $M_\odot$ although the error bars are also consistent with a relatively flat IMF in this
regime.  

It should be noted that the exact shape of the substellar IMF is somewhat dependent on the choice of bin centers and sizes.  For
a bin width of 0.3 dex, shifting the bin centers in increments of 0.05 dex shifts the location of both the primary and secondary
peaks through a range of masses from $\sim$0.25-0.1 $M_\odot$ and $\sim$0.03-0.04 $M_\odot$ respectively.  Additionally, in some
cases the secondary peak disappears and the substellar IMF becomes flat.  Decreasing the bin width by 30\% emphasizes the
secondary peak, however, the errors remain consistent with a flat IMF. Increasing the bin widths by 30\% either preserves the
secondary peak, flattens the substellar mass function, or causes it to decline throughout the brown dwarf regime depending on the
choice of bin centers.

\section{DISCUSSION}

The low mass IMF has been investigated for a number of other young star forming regions and in some cases the results are strikingly
similar to the IMF derived here.  For example, a variety of photometric and spectroscopic surveys have been completed in the extremely
dense Trapezium cluster and the intermediate density IC 348 cluster by multiple authors
\citep{hc00,luhman00,muench02,luhman03b,muench03,sles04,lucas05} and all find mass functions rising to a maximum at $\sim$0.1-0.2
$M_\odot$ before declining into the brown dwarf regime, consistent with our NGC 2024 IMF peak at $\sim$0.2 $M_\odot$.  These results are
very different from the mass function derived for the Taurus star forming region (characterized by its low gas and stellar densities)
which peaks at 0.8 $M_\odot$ \citep{briceno02,luhman03a,luhman04}.  In addition, \citet{muench02} and \citet{sles04} cite a possible
secondary peak in the Trapezium IMF below the hydrogen-burning limit although the locations of their respective peaks differ both from
each other and our possible secondary peak ($\sim$0.05 $M_\odot$ for Slesnick et al., $\sim$0.025 $M_\odot$ for Muench et al., as
compared to $\sim$0.035 $M_\odot$ for NGC 2024).  If this peak is a real feature in the mass function  it may indicate a break in
formation mechanism for low mass objects \citep{muench02}.  On the other hand, it may be an artifact introduced by the mass-luminosity
relation for brown dwarfs and not a true reflection of the mass function itself \citep{muench03}.  In addition, as discussed in \S
\ref{imf}, in the case of NGC 2024 the exact shape of the substellar mass function is influenced by binning thus the secondary peak may
be a consequence of small number statistics and incompleteness below 0.04 $M_\odot$.  

A more robust tool for investigating the differences between low mass IMFs is the ratio of brown dwarfs to stars as
this quantity is independent of the detailed structure and exact shape of cluster mass functions. 
\citet{briceno02} define the ratio of the numbers of stellar and substellar objects as  \begin{equation}
R_{ss}=\frac{N(0.02\le M/M_\odot \le 0.08)}{N(0.08 < M/M_\odot \le 10)}.  \end{equation}  In our completeness
corrected, extinction limited mass function for NGC 2024 there are 45 objects with masses
$0.02M_\odot$$<$$M$$<$$0.08 M_\odot$ and 103 objects with masses $M>0.08 M_\odot$.  In addition, there are 36
sources in our photometric catalog with $K$ magnitudes brighter than the bright limit of our mass function
($K$=11.25, \S\ref{imf}). Since the youngest (and thus brightest) object classified as substellar has a $K$
magnitude of 11.75 with the majority of brown dwarfs falling below $K$=12.75 it is reasonable to infer that all of
these bright photometric sources have masses greater than 0.08 $M_\odot$.  Finally, we include 9 sources from the
2MASS catalog with magnitudes brighter than the FLAMINGOS saturation limit which are also expected to be far more
massive than the substellar limit.  This yields a value of $R_{ss}$=45/148 or 0.30$\pm$0.05 assuming Poisson
errors.  

The ratio of brown dwarfs to stars has been derived for a number of young star forming regions.  In the Trapezium,
\citet{sles04} find $R_{ss}$=0.20 which is slightly lower than the value of $R_{ss}$=0.26$\pm$0.04 found by \citet{luhman00}. 
This variation in $R_{ss}$ is likely caused by the use of different evolutionary models.  \citet{sles04} employed DM97 tracks
to derive masses whereas \citet{luhman00} used the BCAH98 models.   We have recomputed the Trapezium $R_{ss}$ using the data
of \citet{sles04} with the BCAH98 models and find a higher value of $R_{ss}$=0.30, in excellent agreement with our result for
NGC 2024.  Similarly, looking at Figure \ref{hrdiagrams} we note that had we chosen to employ the DM97 models we would have
counted fewer brown dwarfs, leading to a lower value of $R_{ss}$ in closer agreement with the result of \citet{sles04}.  To
remain consistent in the ensuing discussion, however, from this point forward we will discuss only those values derived using
the BCAH98 models. 

\citet{luhman03b} use the BCAH98 models to compute $R_{ss}$ for both the Taurus aggregates and the IC 348
cluster and find significantly lower $R_{ss}$ values: $R_{ss}$=0.14$\pm$0.04 for Taurus and
$R_{ss}$=0.12$\pm$0.03 for IC348.   However, as surveys of Taurus have been expanded beyond the aggregate
radii, the corresponding ratio of brown dwarfs to stars has also increased. \citet{luhman04} combined optical
imaging and spectroscopy with infrared photometry from 2MASS to find $R_{ss}$=0.18$\pm$0.04) for a 12.4 deg$^2$
area.  More recently, \citet{guieu06} completed an optical survey covering $\sim$28 deg$^2$, and find an
updated $R_{ss}$ for Taurus of 0.24$\pm$0.05. These results apparently bring the region into agreement with the
Trapezium and (within the 1 sigma errors) NGC 2024, possibly indicating that brown dwarf formation occurs
independent of the local environment.  On the other hand, both sets of authors note that their Taurus surveys
may suffer from possible incompleteness at low stellar masses (M=0.3-0.6M$_\odot$, spectral types
$\simeq$M2-M4). Further, Taurus is a very extendend region ($>$ 100 deg$^2$) and both surveys, while
significant do not yet have complete spatial coverage of the distributed population.  Detections of additional
low stellar mass objects by forthcoming surveys of Taurus may act to drive the $R_{ss}$ values back down.

\citet{guieu06} also examined the dependence of $R_{ss}$ on radial distance from the Taurus aggregates.  They find that the number of
brown dwarfs in the aggregate centers ($r<0.7R_{agg}$, where $R_{agg}$ is typically 0.5-1.1pc) is depeleted by a factor of 2.3
compared to the more extended population ($r\geq$$R_{agg}$).  \citet{guieu06} conclude that this is best explained if brown dwarf
formation is dominated by the embryo-ejection model \citep{reipurthclarke01} and the new brown dwarfs observed in the distributed
Taurus population are ejected unbound objects that have dynamically evolved away from their birth sites.   Unfortunately this model
cannot explain the low value of $R_{ss}$ in IC 348. According to the simulations of \citet{kb03}, the distance traveled by an ejected
brown dwarf in time $\tau$ is inversely dependent on the size of the gravitational potential from which it escapes.   The mass
contained in the core of IC 348 ($\sim$100 $M_\odot$ within 0.5 pc, \citealp{ll95}) is significantly larger than the mass of a typical
Taurus aggregate (M$<50M_\odot$), however, their ages are similar (IC 348 and the distributed population of Taurus are both estimated
to have $\tau\sim$2 myr).  Consequently we can expect the ejected brown dwarf population in IC 348 to be contained within a radius
smaller than the ejected population in Taurus.  As mentioned above, $R_{ss}$ in Taurus converges to 0.24 within 0.5-1.1 pc. 
\citet{muench03} investigated the radial dependence of the brown dwarf fraction in IC 348 on scales 0.5-1 pc and find no increase in
the number of brown dwarfs at large radii.  In addition, the models of \citet{kb03} predict that there can be {\it at most}
0.25$\pm$0.04 brown dwarfs per star produced as ejected embryos. While formally consistent, this value remains slighly below the
$R_{ss}$ we find for NGC 2024, indicating a possible excess of brown dwarfs in the region.  Clearly, a universal application of the
embryo-ejection model as applied to Taurus cannot simultaneously explain the observed deficit of brown dwarfs in IC 348 and the
possible excess of these objects in NGC 2024.

Another possibility is that the outcome of the brown dwarf formation process is not universal but depends on star forming envronment. 
\citet{briceno02} and \citet{luhman03b} offer the explanation that the differences between the peak mass of the Taurus IMF and the young
cluster IMFs may reflect a disparity among the local Jeans masses.  There is some numerical evidence in support of this theory.  Recent
simulations of turbulent fragmentation by \citet{padoan02,padoan04} and \citet{bate05} find that while the mass distribution for star
forming cores appears to be independent of environment for masses larger than a solar mass, the shape of the mass function for subsolar
masses is dependent on the gas density and local velocity dispersion.  Specifically, for higher gas densities and velocity dispersions,
the IMF peaks at lower masses and produces larger numbers of brown dwarfs.  These results are in agreement with the higher IMF peak for
Taurus as the average gas density in the region has been estimated to be orders of magnitude lower than that of IC 348, the Trapezium, or
NGC 2024.  However, simulations of turbulent fragmentation do not quite account for the excess of brown dwarfs observed in the Trapezium
and NGC 2024 ($R_{ss}$=0.26 and $R_{ss}$=0.30) $R_{ss}$ as compared to IC 348  ($R_{ss}$=0.12). 

An alternate formation mechanism has been put forth by a number of authors \citep[e.g.][]{kb03,wz04,robberto04} who suggest that
isolated brown dwarfs in young clusters can form when the accretion disk around a prestellar core is prematurely eroded by the
strong ionizing radiation emitted from O and B stars.  The ionizing source for NGC 2024 has recently been identified as IRS2b, a
late O to early B main sequence star located in the core of the cluster (cf. Figure \ref{bddist}) and at the center of the radio
continuum radiation field \citep{bik03}.  The presence of an O star in NGC 2024, the detection of low accretion rates in the
Trapezium  (also in the presence of known O and B stars) \citep{robberto04}, and the absence of photoionizing stars in IC 348
indicates that photoevaporation of accretion disks may be a good explanation for the enhanced numbers of brown dwarfs in the
Trapezium and NGC 2024.  If this is the case, we would expect to observe a higher density of brown dwarfs in regions closest to
the ionizing source(s).  Unfortunately we do not have enough statistics in our current data set to test this theory, thus we defer
further discussion to future work and simply note that while the IMF for solar to high mass stars appears uniform, the low mass end may
depend on the local environment, particularly in young clusters where the preferred mechanism for brown dwarf formation may vary
from region to region.

\section{SUMMARY} 

We present FLAMINGOS photometry and spectroscopy for 71 low mass objects in NGC 2024.  Water absorption features in low resolution $J$ and
$H$ band spectra were used to find spectral types ranging from $\sim$M1 to $>$M8 with typical errors in the classifications of 0.5-1
subclasses. A qualitative surface gravity analysis was used to distinguish cluster members from possible background giants.  Spectral
types for the 67 cluster members with M-type spectra were then converted to effective temperatures and photometry was used to calculate
extinctions and bolometric luminosities for each source. Finally, objects were placed on H-R diagrams and masses and ages were inferred
with the assistance of pre-main sequence evolutionary models.  

We find a median age for NGC 2024 of 0.5 Myr using the evolutionary models of \citet{dm97}, which is consistent with
a median age $<$1 Myr as derived from the models of \citet{bcah98}.  Derived masses range from 0.02 $M_\odot$ to 0.72
$M_\odot$ using the tracks of \citet{bcah98}, with 23 of the 67 objects falling below the stellar/substellar
boundary.  When looking at the spatial distribution of these objects, we find that rather than being relegated to one
area of the cluster they appear to be evenly distributed relative to their stellar counterparts.  Thirty nine percent
of our classified brown dwarfs appear to have an infrared excess, possibly indicative of thermal emission from a warm
disk. We define an extinction limited subsample of our spectroscopic sources and combine it with photometry to
construct a mass function for the region.  The IMF for NGC 2024 peaks at $\sim$0.2 $M_\odot$, consistent with peak
values found for the Trapezium and IC 348.  The IMF then declines into the brown dwarf regime but exhibits a possible
secondary peak around 0.035 $M_\odot$. Finally, using this IMF we estimate the ratio of stellar to substellar objects
in NGC 2024 to be $R_{ss}$=0.30$\pm$0.05.  This is consistent with, although slightly higher than the $R_{ss}$
values found for the Trapezium cluster and the distributed Taurus population but roughly a factor of 2 higher than
the $R_{ss}$ found for IC 348. Taken together, these results may imply that the low mass end of the initial mass
function is not universal but rather depends on the star formation environment.

We thank August Muench for many fruitful discussions and we are grateful to Kevin Luhman and the anonymous referee for their helpful
comments on earlier versions of this manuscript.  FLAMINGOS was designed and constructed by the IR instrumentation group (PI: R. Elston)
at the University of Florida, Department of Astronomy with support from NSF grant (AST97-31180) and Kitt Peak National Observatory.  The
data presented in this work were collected under the NOAO Survey Program, "Towards a Complete Near-Infrared Spectroscopic Survey of Giant
Molecular Clouds" (PI: E. Lada) and supported by NSF grants, AST97-3367 and AST02-02976 to the University of Florida. This publication
makes use of data products from the Two Micron All Sky Survey, which is a joint project of the University of Massachusetts and the
Infrared Processing and Analysis Center/California Institute of Technology, funded by the National Aeronautics and Space Administration
and the National Science Foundation.

\clearpage

\begin{deluxetable}{ccccc}
\tablewidth{0pt}
\tablecaption{FLAMINGOS Spectroscopic Observations}
\tablehead{
\colhead{Mask ID} & \colhead{Telescope} & \colhead{Observation Date} & \colhead{t$_{exp}$} & \colhead{N$_{exp}$} }
\startdata
n2024bd1 & 4m & 2003 Jan 19 &   600s & 8  \\*  
oc24mf11 & 4m & 2003 Dec 06 &   300s & 8  \\*  
oc24mf21 & 4m & 2003 Dec 10 &   300s & 13  \\*  
n2024b2 &  4m & 2004 Dec 01 &   300s & 10  \\*  
n2024b3 &  4m & 2004 Dec 01 &   300s & 10  \\*  
n2024f31 & 2.1m & 2003 Nov 29 & 300s & 14 \\*
\enddata
\label{obstable}
\end{deluxetable}
\clearpage

\begin{deluxetable}{lcccc}
\tablewidth{0pt}
\tablecaption{Young Spectral Standards}
\tablehead{
\colhead{ID} & \colhead{R.A.} & \colhead{Dec.} & \colhead{Optical Spectral Type} }
\startdata
I348-052  & 03:44:43.53 & +32:07:43.0  &   M1	     \\*
I348-122  & 03:44:33.22 & +32:15:29.1  &   M2.25    \\* 
I348-207  & 03:44:30.30 & +32:07:42.6  &   M3.5     \\* 
I348-095  & 03:44:21.91 & +32:12:11.6  &   M4	     \\*
I348-266  & 03:44:18.26 & +32:07:32.5  &   M4.75    \\* 
I348-230  & 03:44:35.52 & +32:08:04.5  &   M5.25    \\* 
I348-298  & 03:44:38.88 & +32:06:36.4  &   M6	     \\*
I348-329  & 03:44:15.58 & +32:09:21.9  &   M7.5     \\* 
I348-405  & 03:44:21.15 & +32:06:16.6  &   M8	     \\*
I348-603  & 03:44:33.42 & +32:10:31.4  &   M8.5    \\*  
KPNO-Tau4 & 04:27:28.01 & +26:12:05.3  &   M9.5     \\*
\enddata

\tablecomments{
Spectral types for the IC 348 objects are the spectral types adopted by \citet{luhman03b} and found in
table 2 of that work.  The spectral type for KPNO-Tau4 was determined by \citet{briceno02}. }
\label{standtable}
\end{deluxetable}
\clearpage

\begin{deluxetable}{ccccccrrccc}
\tabletypesize{\scriptsize}
\tablewidth{0pt}
\tablecaption{Measured Data and Derived Quantities for Classified Sources in NGC 2024}
\tablehead{
\colhead{Source} & \colhead{FLAMINGOS ID} & \colhead{J} & \colhead{H} &
\colhead{K} & \colhead{M Subclass\tablenotemark{a}} & \colhead{$A_V$}  & \colhead{r$_K$} &  \colhead{logT$_{eff}$} &
\colhead{log(L/L$_\odot$)} & \colhead{Mass ($M_\odot$)} }
\startdata
01 & FLMN\_J0541378-0153112 & 12.06		     & 11.07 & 10.30  & $<$M  &    ...  &    ...  &    ... &	 ...  &   ...  \\
02 & FLMN\_J0541344-0154409 & 12.25		     & 11.07 & 10.37  & 1.75  &   4.73  &   0.19  &  3.558 &   0.125  &  0.72  \\
03 & FLMN\_J0541366-0154082 & 13.71		     & 11.91 & 11.07  & 4.00  &  10.82  &  -0.11  &  3.514 &  -0.012  &  0.29  \\
04 & FLMN\_J0541402-0150522 & 14.11		     & 12.20 & 11.29  & 1.75  &  11.36  &  -0.03  &  3.558 &  -0.004  &  0.72  \\
05 & FLMN\_J0541373-0151403 & 15.54		     & 12.81 & 11.29  & 1.75  &  18.82  &   0.09  &  3.558 &   0.264  &  0.72  \\
06 & FLMN\_J0541333-0151270 & 13.64		     & 12.00 & 11.37  & 1.75  &   8.91  &  -0.13  &  3.558 &  -0.124  &  0.70  \\
07 & FLMN\_J0541293-0151304 & 13.22		     & 11.96 & 11.43  & 2.50  &   5.59  &  -0.05  &  3.544 &  -0.292  &  0.55  \\
08 & FLMN\_J0541456-0151229 & 14.64		     & 12.69 & 11.65  & 3.50  &  12.18  &  -0.01  &  3.524 &  -0.195  &  0.29  \\
09 & FLMN\_J0541436-0151353 & 15.30		     & 12.80 & 11.66  & 4.00  &  17.18  &  -0.20  &  3.514 &  -0.019  &  0.29  \\
10 & FLMN\_J0541390-0151453 & 15.10		     & 12.87 & 11.69  & 4.75  &  14.70  &  -0.06  &  3.499 &  -0.146  &  0.17  \\
11 & FLMN\_J0541443-0155247 & 15.29		     & 13.01 & 11.69  & 2.75  &  14.93  &   0.12  &  3.539 &  -0.067  &  0.54  \\
12 & FLMN\_J0541417-0151457 & 14.06		     & 12.49 & 11.75  & 4.50  &   8.64  &  -0.09  &  3.504 &  -0.380  &  0.20  \\
13 & FLMN\_J0541496-0153270 & 14.77		     & 12.83 & 11.79  & 2.75  &  11.84  &   0.04  &  3.539 &  -0.218  &  0.54  \\
14 & FLMN\_J0541578-0151278 & 13.57		     & 12.24 & 11.81  & 1.50  &   6.09  &  -0.14  &  3.562 &  -0.393  &  0.68  \\
15 & FLMN\_J0541569-0150331 & 14.18		     & 12.55 & 11.84  & 2.75  &   9.32  &  -0.12  &  3.539 &  -0.350  &  0.44  \\
16 & FLMN\_J0541566-0155526 & 13.58		     & 12.38 & 11.92  & 5.25  &   5.48  &  -0.17  &  3.488 &  -0.587  &  0.13  \\
17 & FLMN\_J0541259-0150243 & 14.36\tablenotemark{b} & 12.94 & 12.20  & 5.25  &   7.48  &  -0.05  &  3.488 &  -0.627  &  0.13  \\
18 & FLMN\_J0541313-0154052 & 14.13		     & 12.78 & 12.23  & 2.75  &   6.73  &  -0.13  &  3.539 &  -0.623  &  0.28  \\
19 & FLMN\_J0541350-0153286 & 14.42		     & 13.08 & 12.41  & 4.75  &   6.61  &  -0.04  &  3.499 &  -0.725  &  0.17  \\
20 & FLMN\_J0541316-0152318 & 14.42		     & 13.13 & 12.49  & 5.50  &   6.36  &  -0.08  &  3.483 &  -0.791  &  0.12  \\
21 & FLMN\_J0541438-0151399 & 17.57		     & 14.30 & 12.52  & 4.50  &  24.09  &  -0.06  &  3.504 &  -0.131  &  0.20  \\
22 & FLMN\_J0541489-0152297 & 16.19		     & 13.82 & 12.65  & $<$M  &   ...   &   ...   &   ...  &	...   &   ...  \\
23 & FLMN\_J0541350-0155300 & 16.81		     & 14.37 & 12.67  & 5.50  &  16.82  &   0.30  &  3.483 &  -0.487  &  0.12  \\
24 & FLMN\_J0541501-0155320 & 18.04		     & 14.65 & 12.73  & 4.75  &  25.25  &  -0.01  &  3.499 &  -0.182  &  0.17  \\
25 & FLMN\_J0541506-0158041 & 14.33		     & 13.36 & 12.77  & 7.75  &   3.49  &  -0.05  &  3.431 &  -1.084  &  0.03  \\
26 & FLMN\_J0541379-0153187 & 16.18		     & 14.20 & 12.94  & 4.75  &  12.43  &   0.16  &  3.499 &  -0.727  &  0.17  \\
27 & FLMN\_J0541460-0150071 & 14.91		     & 13.64 & 12.98  & 5.00  &   6.05  &  -0.03  &  3.494 &  -0.982  &  0.17  \\
28 & FLMN\_J0541364-0154412 & 15.60		     & 13.93 & 12.98  & 7.25  &  10.25  &  -0.10  &  3.443 &  -0.907  &  0.04  \\
29 & FLMN\_J0541456-0157332 & 14.25		     & 13.45 & 12.99  & 8.00  &   1.15  &  -0.05  &  3.425 &  -1.264  &  0.03  \\
30 & FLMN\_J0541314-0154347 & 16.15		     & 14.24 & 13.00  & 3.25  &  11.86  &   0.24  &  3.529 &  -0.723  &  0.35  \\
31 & FLMN\_J0541469-0151099 & 15.55		     & 13.84 & 13.01  & 4.25  &   9.95  &  -0.08  &  3.509 &  -0.828  &  0.25  \\
32 & FLMN\_J0541514-0152470 & 15.59		     & 13.96 & 13.03  & 5.75  &   9.59  &  -0.02  &  3.478 &  -0.900  &  0.10  \\
33 & FLMN\_J0541588-0150311 & 15.12		     & 13.84 & 13.10  & 5.25  &   6.20  &   0.03  &  3.488 &  -1.033  &  0.15  \\
34 & FLMN\_J0541413-0150300 & 16.33		     & 14.28 & 13.12  & 5.00  &  13.14  &   0.01  &  3.494 &  -0.783  &  0.15  \\
35 & FLMN\_J0541281-0152138 & 15.10		     & 13.85 & 13.19  & 5.25  &   5.93  &  -0.03  &  3.488 &  -1.078  &  0.15  \\
36 & FLMN\_J0541359-0153371 & 15.57		     & 14.07 & 13.22  & 7.25  &   8.70  &  -0.10  &  3.443 &  -1.059  &  0.04  \\
37 & FLMN\_J0541480-0152198 & 16.03		     & 14.22 & 13.25  & 5.25  &  11.02  &  -0.05  &  3.488 &  -0.919  &  0.13  \\
38 & FLMN\_J0541439-0153513 & 16.83		     & 14.84 & 13.28  & 3.00  &  12.36  &   0.62  &  3.534 &  -0.810  &  0.37  \\
39 & FLMN\_J0541461-0151442 & 18.86		     & 15.30 & 13.37  & 7.00  &  27.05  &  -0.18  &  3.449 &  -0.450  &  0.04  \\
40 & FLMN\_J0541345-0152135 & 15.55		     & 14.20 & 13.37  & 8.00  &   6.16  &  -0.01  &  3.425 &  -1.236  &  0.03  \\
41 & FLMN\_J0542029-0152038 & 16.13		     & 14.45 & 13.42  & 5.50  &   9.91  &   0.06  &  3.483 &  -1.036  &  0.12  \\
42 & FLMN\_J0541390-0155008 & 17.66		     & 14.98 & 13.43  & 6.00  &  19.27  &  -0.04  &  3.472 &  -0.720  &  0.09  \\
43 & FLMN\_J0541528-0150162 & 15.71		     & 14.33 & 13.50  & 7.50  &   8.00  &  -0.09  &  3.437 &  -1.205  &  0.03  \\
44 & FLMN\_J0541387-0150507 & 16.11		     & 14.54 & 13.56  & 6.75  &   8.98  &   0.02  &  3.455 &  -1.168  &  0.05  \\
45 & FLMN\_J0541508-0152236 & 15.77		     & 14.38 & 13.57  & 5.50  &   7.27  &   0.02  &  3.483 &  -1.191  &  0.13  \\
46 & FLMN\_J0541523-0157123 & 15.75		     & 14.35 & 13.57  & 5.75  &   7.50  &  -0.03  &  3.478 &  -1.191  &  0.11  \\
47 & FLMN\_J0541475-0150052 & 15.96\tablenotemark{b} & 14.47 & 13.62  & 4.75  &   7.98  &   0.04  &  3.499 &  -1.160  &  0.19  \\
48 & FLMN\_J0541377-0155097 & 16.99		     & 14.93 & 13.65  & 7.75  &  13.40  &  -0.01  &  3.431 &  -1.079  &  0.03  \\
49 & FLMN\_J0541514-0152408 & 17.02		     & 14.97 & 13.67  & 6.50  &  13.36  &   0.07  &  3.461 &  -1.045  &  0.06  \\
50 & FLMN\_J0541372-0151580 & 17.28		     & 14.96 & 13.69  & 4.75  &  15.52  &  -0.03  &  3.499 &  -0.916  &  0.20  \\
51 & FLMN\_J0541534-0156482 & 15.60		     & 14.35 & 13.70  & 4.00  &   5.82  &   0.01  &  3.514 &  -1.244  &  0.23  \\
52 & FLMN\_J0541537-0157503 & 16.30		     & 14.73 & 13.70  & 6.50  &   9.00  &   0.08  &  3.461 &  -1.215  &  0.06  \\
53 & FLMN\_J0541506-0153064 & 16.54		     & 14.86 & 13.71  & 8.25  &   9.12  &   0.10  &  3.419 &  -1.274  &  0.03  \\
54 & FLMN\_J0541340-0149391 & 18.05		     & 15.36 & 13.75  & 4.25  &  18.86  &   0.11  &  3.509 &  -0.803  &  0.24  \\
55 & FLMN\_J0541497-0156198 & ...		     & 15.88 & 13.76  & 7.00  &  29.11  &  -0.14  &  3.449 &  -0.531  &  0.04  \\
56 & FLMN\_J0541374-0152311 & 16.40		     & 14.77 & 13.76  & 6.75  &   9.52  &   0.02  &  3.455 &  -1.228  &  0.05  \\
57 & FLMN\_J0541328-0151271 & 15.91		     & 14.64 & 13.82  & 8.50  &   5.36  &   0.01  &  3.412 &  -1.461  &  0.02  \\
58 & FLMN\_J0541366-0154151 & 17.24		     & 15.32 & 14.00  & 7.50  &  12.91  &   0.07  &  3.437 &  -1.228  &  0.03  \\
59 & FLMN\_J0541501-0151180 & 16.13		     & 14.86 & 14.02  & 7.50  &   7.00  &  -0.02  &  3.437 &  -1.449  &  0.03  \\
60\tablenotemark{d} & FLMN\_J0541390-0151316 & 17.86 & 15.64 & 14.19  & 4.00  &  14.64  &   0.24  &  3.514 &  -1.123  &  0.24  \\
61 & FLMN\_J0541357-0153483 & 16.40		     & 15.03 & 14.21  & 2.75  &   6.66  &   0.16  &  3.539 &  -1.373  &  0.37  \\
62 & FLMN\_J0541287-0155264 & 16.44		     & 15.02 & 14.24  & 7.25  &   7.98  &  -0.12  &  3.443 &  -1.493  &  0.04  \\
63 & FLMN\_J0541380-0151536 & 17.56		     & 15.63 & 14.37  & 4.75  &  11.98  &   0.19  &  3.499 &  -1.316  &  0.16  \\
64\tablenotemark{d} & FLMN\_J0541483-0151304 & 17.44 & 15.60 & 14.42  & 5.00  &  11.23  &   0.15  &  3.494 &  -1.371  &  0.14  \\
65 & FLMN\_J0542020-0153542 & 16.68		     & 15.31 & 14.50  & 6.25  &   7.27  &  -0.01  &  3.466 &  -1.588  &  0.07  \\
66 & FLMN\_J0541345-0154500 & 17.10		     & 15.54 & 14.61  & 7.75  &   8.85  &  -0.06  &  3.431 &  -1.626  &  0.03  \\
67 & FLMN\_J0541547-0151461 & 17.74		     & 16.11 & 15.05  & 8.00  &   8.70  &   0.05  &  3.425 &  -1.816  &  0.03  \\
68 & FLMN\_J0541250-0152267 & 13.74\tablenotemark{b} & 12.25 & 11.59  & 5.00\tablenotemark{c}  &   8.05  &  -0.14  & 3.494  & -0.399 &  0.15\\
69 & FLMN\_J0541338-0153087 & 13.83		     & 12.48 & 11.75  & 6.75\tablenotemark{c}  &   6.98  &  -0.09  & 3.455  & -0.516 &  0.05 \\
70 & FLMN\_J0541269-0154093 & 13.88		     & 12.72 & 12.04  & 2.50\tablenotemark{c}  &   4.68  &   0.16  & 3.544  & -0.568 &  0.47 \\
71 & FLMN\_J0541368-0154479 & 15.54		     & 13.43 & 12.07  & 7.25\tablenotemark{c}  &  14.25  &   0.03  & 3.443  & -0.399 &  0.04 \\
\enddata
\tablenotetext{a}{Spectral types are listed as M subclasses, thus a table entry of 0.0=M0.0, 7.50=M7.50, etc.}
\tablenotetext{b}{Sources 17, 47, and 68 have J magnitudes derived from FLAMINGOS imaging on the 2.1m telescope.}
\tablenotetext{c}{Sources 68-71 have spectral types derived from 2.1m spectroscopy.}
\tablenotetext{d}{Possible background giant.}
\label{datatable}
\end{deluxetable}
\clearpage
\clearpage

\clearpage

\clearpage

\clearpage

\clearpage

\clearpage

\clearpage

\clearpage

\clearpage

\end{document}